\documentclass{aa}
\usepackage{graphicx,natbib,bm,url,color}
\graphicspath{{./fig/}}
\usepackage{hyperref}
\usepackage{macros}

\begin{document}

\title{Isochrones in primordial magnetic field evolution}

\author{A. Brandenburg\inst{1,2,3,4}
\and M. Cielo\inst{1}
\and O. Iarygina\inst{1,5,6}
\and F. Vazza\inst{7}
}
\institute{
Nordita, KTH Royal Institute of Technology and Stockholm University, Hannes Alfv\'ens v\"ag 12, 10691 Stockholm, Sweden, \email{brandenb@nordita.org}
\and
The Oskar Klein Centre, Department of Astronomy, Stockholm University, AlbaNova, 10691 Stockholm, Sweden
\and
McWilliams Center for Cosmology \& Department of Physics, Carnegie Mellon University, 5000 Forbes Ave, Pittsburgh, PA 15213, USA
\and
School of Natural Sciences and Medicine, Ilia State University, 3-5 Cholokashvili Avenue, 0194 Tbilisi, Georgia
\and
The Oskar Klein Centre, Department of Physics, Stockholm University, AlbaNova, 10691
Stockholm, Sweden
\and
Institute Lorentz of Theoretical Physics, Leiden University, 2333 CA Leiden, The Netherlands
\and
Dipartimento di Fisica e Astronomia, Universita di Bologna, Via Gobetti 93/2, 40129 Bologna, Italy}

\date{}

\abstract
{
In the early universe, a primordial magnetic field undergoes a turbulent decay
while its length scale increases due to an inverse cascade.
The size of the largest processed eddy scales with the Alfv\'en speed and grows with time.
In a diagram of Alfv\'en speed vs.\ length scale, all possible solutions must lie on a line through the origin with a slope proportional to the inverse of the present time.
In principle, however, such lines can also be defined for earlier times.
}{
The lines for earlier times form isochrones that may be observationally accessible, for example through the magnetically driven stochastic gravitational wave background.
However, the position and slope of these isochrones is sensitive to the zero point of the time.
Here, we show that for any initial magnetic field, a proper time can be determined such that the resulting isochrones
at early times are nearly parallel to those at late times, i.e., they have the same slope.
}{
We use two-dimensional numerical simulations of decaying MHD turbulence and vary the initial position of the peak of the magnetic energy spectrum.
In this case, the evolution is governed by the conservation of anastrophy.
}{
A fit to the Alfv\'en time yields an accurate estimate of the factor by which the decay time
is longer than the Alfv\'en time, while the offset in the fit provides an estimate of the proper time
that needs to be added to the nominal time since the beginning of each simulation.
We also find that the presence of an initial velocity field of realistic strength helps producing a more straight track from the beginning.
}{
The magnetic field parameters lie on universal isochrones even for early times.
They provide a testable framework for magnetic fields generated at times as early as the end of inflation,
starting with the time of reheating.
}


\maketitle
\nolinenumbers

\section{Introduction}

Primordial magnetic fields may still be observable in the voids of the cosmic web, while the pollution of magnetic fields by purely 
 astrophysical (i.e., galaxy evolution-driven using expulsion of magnetized winds and/or jets) 
mechanisms are believed to be incapable of producing large enough filling factors of the magnetic field,
as seen in numerical simulations \citep[e.g.,][]{2010MNRAS.401...47D,2022A&A...660A..80B,tj24,Vazza2025,2026PhRvD.113b3523G}. 
At the simplest level, primordial magnetic fields are characterized by their typical strengths and length scales.
A useful measure of their rms value $\Brms$ is the Alfv\'en speed $\vA=\Brms/\sqrt{\mu_0\rho}$,
with $\rho$ being the density and $\mu_0$ the vacuum permeability: it quantifies the speed at which a magnetic field perturbation spreads.
Thus, after a certain time $t$, the typical radius of such a perturbation is $\sim \vA t$.

Primordial magnetic fields are generated on scales smaller than the horizon scale at the epoch of generation \citep[e.g.,][]{GR01}.
Yet, if the generation happens during or before inflation the maximum scales can be larger than the present-day horizon
\citep[e.g.,][]{GR01,2008PhRvD..77d3529D,wi11,2013A&ARv..21...62D,sub15}.
Of course, the Universe expands, but this expansion is commonly scaled out and so one talks about
comoving length scales and comoving magnetic field strengths to account for the dilution of the field with time.
For example, a physical magnetic field of $10^{24}\G$ at the electroweak epoch corresponds to a comoving field of $1\uG$ today,
and the horizon scale of about $1\cm$ corresponds to $6\times10^{-4}\pc$ \citep{Kahn+13}.
But the actual magnetic field is weaker still because of turbulence, which speeds up the decay of the field.
However, as the magnetic field undergoes a turbulently enhanced decay, it can still grow on large length scales \citep{BEO96}.
Thus, there is an increase of the magnetic energy spectrum at small wavenumbers.
This is due to what is called an inverse cascade \citep{Frisch+75,PFL76}.

The speed at which the turbulent eddies grow is expected to be limited by $\vA$.
In the literature, one talks about the ``largest processed eddy'' \citep{Kahn+13}, whose size is $\sim \vA t$.
This leads us to the expectation that in our Universe, whose
age is $t_\mathrm{U}\approx 14\Gyr$, there should be no causally generated magnetic fields whose characteristic length scale is larger than $\vA t_\mathrm{U}$.

Using $\rho=4\times10^{-31}\g\cm^{-3}$ for the baryon density and $\Brms=1\nG$ for the magnetic field,
we find $\vA=4\km\s^{-1}$ and therefore the largest processed eddy has a size of $60\kpc$.
For a weaker field, $\vA t$ will be correspondingly smaller.
In a diagnostic diagram of magnetic field strengths or $\vA$ vs.\ characteristic length scale $\xiM$, the above considerations lead
to a line $\vA=\xiM(t_\mathrm{U})/t_\mathrm{U}$, below which no causally generated magnetic fields should ever be detected within the time $t_\mathrm{U}$ \citep{BJ04}.

The existence of a line $\vA=\xiM(t_\mathrm{U})/t_\mathrm{U}$ is actually quite remarkable,
because it means that all observed combinations of $\vA$ and $\xiM$ must lie on this line,
which is independent of the initial conditions.
Even for very weak magnetic fields, the present value of $\vA$ would still lie on this line,
but the point would just be shifted further to the lower left along this line.
This remarkable realization of \citep{BJ04} was emphasized by \cite{HS23} who wrote that
``Its great utility, which has perhaps not been spelled out explicitly, is that it implies that one need not know
the functional form ... during the early stages of the decay in order to compute $B$ and $\xiM$ at later times''.

Using numerical simulations and theoretical considerations, we can address
the question of where in the diagnostic diagram the parameters $\vA$ and $\xiM$ of the universe lie at any earlier time.
We thus expect the $\vA$ vs.\ $\xiM$ diagnostic diagram to be occupied by isochrones given through
\begin{equation}
\vA(\xiM)=\xiM(t_\ast)/t_\ast,
\end{equation}
where $t_\ast$ denotes the time of the isochrone.
These isochrones would then all be parallel to each other.

Over the past few decades, many numerical simulations have been performed, where the initial field obeyed a
certain magnetic energy spectrum $\EM(k)$ with a peak at the wavenumber $k=k_B$ and a field strength $B_0$ related to the integral of $\EM(k)$.
This implies that in the $\vA$ vs.\ $\xiM$ diagram, any point would represent a valid initial condition.
This seems to be in conflict with the idea of universal isochrones that are given by $\xiM(t_\ast)/t_\ast$,
because it would restrict permissible initial conditions to lie on the line $\xiM(t_\ast)/t_\ast$ in the $\vA$ vs.\ $\xiM$ diagram.

\cite{HS23} have emphasized that the early stages do not matter for computing $\vA$ and $\xiM$ at later times.
This could address the puzzle about the permissible initial conditions.
However, it would leave open the question of when the theoretical isochrones become parallel to those at late times and how they will be reached in practice.

Can the ideas of the theoretical isochrones be applied already to the moments shortly after the end of inflation, for example?
This question is relevant when one compares different magnetogenesis scenarios, where magnetic fields could be produced over a range of magnetic energy and length scales.
The definition of isochrones is then particularly sensitive to the zero point of the time axis, which is a priori a matter of convention.
In view of the fact that early universe magnetic fields can imprint signatures on the stochastic gravitational wave background
\citep{Caprini+Durrer01, Caprini+Durrer06, Kahniashvili+05, Kahniashvili+08}, the definition of isochrones from such epochs may also be observationally relevant.

In the present paper, we demonstrate that, to compare different magnetogenesis scenarios,
one must first reset the nominal time by a certain amount such that the initial value of $\vA t$ agrees with the initial scale $\xiM$.
This can be done by determining a fit of the instantaneous Alfv\'en time versus $t$, as we discuss in the present paper.

In addition to the uncertainty in the zero point of the time axis, there is a well-known problem with the use of the Alfv\'en time itself.
In fact, it has been found numerically that the actual decay time of MHD turbulence exceeds the Alfv\'en time by a certain
Lundquist number-dependent factor \citep{Bhat+21}, which we call $\CM$.
In that case, the actual isochrones are expected to lie on somewhat higher lines than the Alfv\'enic isochrones; see \cite{HS23}.
The factor $\CM$ was found to depend on the Lundquist number $\Lu$, but saturated at the value of around 50 \citep{BNV24}.
On theoretical grounds, one might expect there to be an additional dependence on the magnetic Prandtl number \citep{HS23}.
This, however, hinges on the validity of earlier findings that the reconnection rate depends on the magnetic Prandtl number \citep{Comisso+15},
which has recently been challenged \citep{KB26}.
In the present paper, we are able to avoid the uncertainty related to the value of $\CM$ as well.
This factor automatically emerges from the aforementioned fit of the instantaneous Alfv\'en time to the nominal time $t$.

Following earlier work \citep{BNV24}, we focus on the 2D case.
It is different from the 3D counterpart in that the governing conserved quantity is here the
anastrophy rather than the mean magnetic helicity density in the helical case or the Hosking integral in the non-helical case.
However, our goal is to clarify the proper definition of the isochrones, and this can equally well be studied in the simpler 2D case.
Furthermore, in 2D, we can reach a larger dynamical range and therefore reduce some artifacts resulting from the finite cutoffs at small and large wavenumbers of the domain.

\section{The model}

\subsection{Basic equations}
\label{BasicEquations}

We perform 2D simulations in Cartesian coordinates, $(x,y,z)$, in a 2D domain of size $L^2$.
The magnetic field is assumed to lie entirely in the $xy$ plane and can therefore be written as $\BB=\nab\times(\zzz A_z)$.
The induction equation then reduces to
\begin{equation}
\frac{\DD A_z}{\DD t}=\eta\nabla^2 A_z,
\label{dAdt2D}
\end{equation}
where $\DD/\DD t=\partial/\partial t+\uu\cdot\nab$ is the advection operator,
$\uu$ is the velocity, and $\eta$ is the magnetic diffusivity.
In the limit $\eta\to0$, \Eq{dAdt2D} obeys the conservation of anastrophy,
$I_\mathrm{A}\equiv\bra{A_z^2}=\const$ \citep{Fyfe+Montgomery76, Pouquet78, Pouquet93},
where angle brackets denote volume averaging.

We also solve the equations for $\uu$ and the logarithmic density $\ln\rho$,
\begin{equation}
\frac{\DD\uu}{\DD t}=
\frac{1}{\rho}\left[\JJ\times\BB+\nab\cdot(2\rho\nu\SSSS)\right]
-\cs^2\nab\ln\rho,
\label{DuDt}
\end{equation}
\begin{equation}
\frac{\DD\ln\rho}{\DD t}=-\nab\cdot\uu.
\label{DlnrhoDt}
\end{equation}
In \Eq{DuDt}, ${\sf S}_{ij}=(\partial_i u_j+\partial_j u_i)/2-\delta_{ij}\nab\cdot\uu/3$
are the components of the rate-of-strain tensor $\SSSS$, $\nu$ is the viscosity,
$\cs$ is the isothermal sound speed, and $\JJ=\nab\times\BB/\mu_0$ is the current density.
We apply periodic boundary conditions in all directions.

\subsection{Initial conditions}
\label{InitialConditions}
Our initial magnetic and velocity fields are given in Fourier space
(indicated by a tilde) by $\tilde{A}_z({\kk})=B_0\, g(\kk) \, S_B(k)$ and $\tilde{\psi}_z({\kk})=u_0\, g(\kk) \, S_u(k)$, respectively,
where $g(\kk)$ is the Fourier transform of a $\delta$-correlated vector field with Gaussian fluctuations,
$\tilde{\psi}_z$ is the Fourier transform of the stream function of the velocity $\uu=\nab\times(\psi_z\zzz)$,
and $k_{v}$ is the initial wavenumber of the energy-carrying eddies for the field $v$,
where $v=B$ or $u$ for the magnetic and velocity fields, respectively.
Likewise, $S_{v}(k)$ with $v=B$ or $u$ determines the spectral shape with \citep{Bra20}
\begin{equation}
S_{v}(k)=\frac{k_{v}^{-d/2} (k/k_{v})^{(\alpha^\mathrm{sub}_{v}+d-1)/2}}
{[1+(k/k_{v})^{2\left(\alpha^\mathrm{sub}_{v}-\alpha^\mathrm{ine}_{v}\right)}]^{1/4}}\;e^{-(k/K_{v})^{n_{v}}},
\label{Sfunction}
\end{equation}
where $d=2$ is the dimension for our 2D runs and $K_{v}$ determines the wavenumber of an exponential cutoff.
In all cases, we use $K_{B}=\infty$ (no cutoff), but for runs with $u_0\neq0$, we use a finite value for $K_{u}$; see below.
Here, $\alpha^\mathrm{sub}_{v}$ is the subinertial range slope and $\alpha^\mathrm{ine}_{v}$ is the inertial range slope.
For a causal spectrum, as it is often used as initial condition for the magnetic field \citep{Durrer+98, CHB01, DC03},
we would use $\alpha^\mathrm{sub}_{v}=3$ in 2D, which results in a $\delta$-correlated magnetic vector potential, but it turns out that after some time,
the slope steepens; see \App{SubinertialSlope}.
Therefore, we use $\alpha^\mathrm{sub}_{v}=4$, which is the value also used in three-dimensional simulations.

It turns out that the spectrum develops an even steeper part $\propto k^5$ near the peak.
Its detailed physical nature has not yet been studied.
We note, however, that this effect is reminiscent of the suspected formation of a ``$k^4$ bulge'' in simulations with an initial $k^2$ spectrum;
see Fig.~6 of the supplemental material of \cite{HS23}.
So far, however, this has not been observed in such a case; see \cite{Reppin+Banerjee17} and \cite{BSV23}
for earlier work on 3D nonhelical turbulence with initial subinertial range spectra between $k^2$ and $k^4$.

In all cases, we use $\Pm\equiv\nu/\eta=1$ for the magnetic Prandtl number.
The density is in all cases initially uniform and equal to $\rho_0$.
The initial magnetic field strength is such that $\vA$ remains subsonic, i.e., $\vA\ll\cs$.
This implies that compressibility effects are weak.
In most of our runs, we used zero velocity initially, i.e., we set $u_0=0$ in 
the expression for $\tilde{\psi}_z({\kk})$ in \Sec{InitialConditions}.
However, in some cases, we want to perform a simulation in a larger domain to cover the inverse cascade at later times.
In this case, we tested two options. First, we start the simulation using a zero velocity initial condition, which however causes
an extra adjustment phase, during which a nearly selfsimilarly evolving velocity field builds up (Run C4).
In a second option and to minimize the artifacts resulting from such an adjustment phase, we start the run by introducing a nonzero initial velocity
with a relative amplitude $u_0/B_0$ of typically around 0.3 (Run C4u); see \Tab{TSummary}, where we give a detailed list of our simulations.

We solve \Eqss{dAdt2D}{DlnrhoDt} using the {\sc Pencil Code} \citep{PC},
where all relevant diagnostics is being computed during run time.
Our numerical resolution is varied between $1024^2$ and $8192^2$ mesh points.
The lowest wavenumber in the domain is given by $k_1=2\pi/L$, where the domain size $L$ is varied such that
the peak wavenumber $k_B$ fits well into the domain.
We use nondimensional quantities by setting $k_B(A)=\cs=\rho_0=\mu_0=1$.
Here, $k_B(A)$ is the magnetic peak wavenumber for our reference point A discussed below.

\subsection{Diagnostic quantities}
\label{DiagnosticQuantities}

We characterize the magnetic field by its mean energy density $\EEM=\bra{\BB^2}/2\mu_0$ and the magnetic integral scale $\xiM$.
Both $\EEM(t)$ and $\xiM(t)$ can be defined in terms of the magnetic energy spectrum $\EM(k,t),$ such that $\EEM=\int\EM\,\dd k$ and $\xiM=\int k^{-1}\EM\,\dd k/\EEM$.
For our reference point A with $k_B(A)=1$, we then have $\xiM\approx1$.

Instead of $\EEM(t)$, we often use $\vA(t)$.
Because of mass conservation and since the runs are only weakly compressible, we have $\rho\approx\rho_0$,
so $\vA\approx\sqrt{2\rho_0\mu_0\EEM(t)}$ is a good approximation.
Note also that the value of $I_\mathrm{A}$ is of the order of $\vA^2\xiM^2$.

\subsection{Simulation strategy}

\begin{figure}[t!]\begin{center}
\includegraphics[width=\columnwidth]{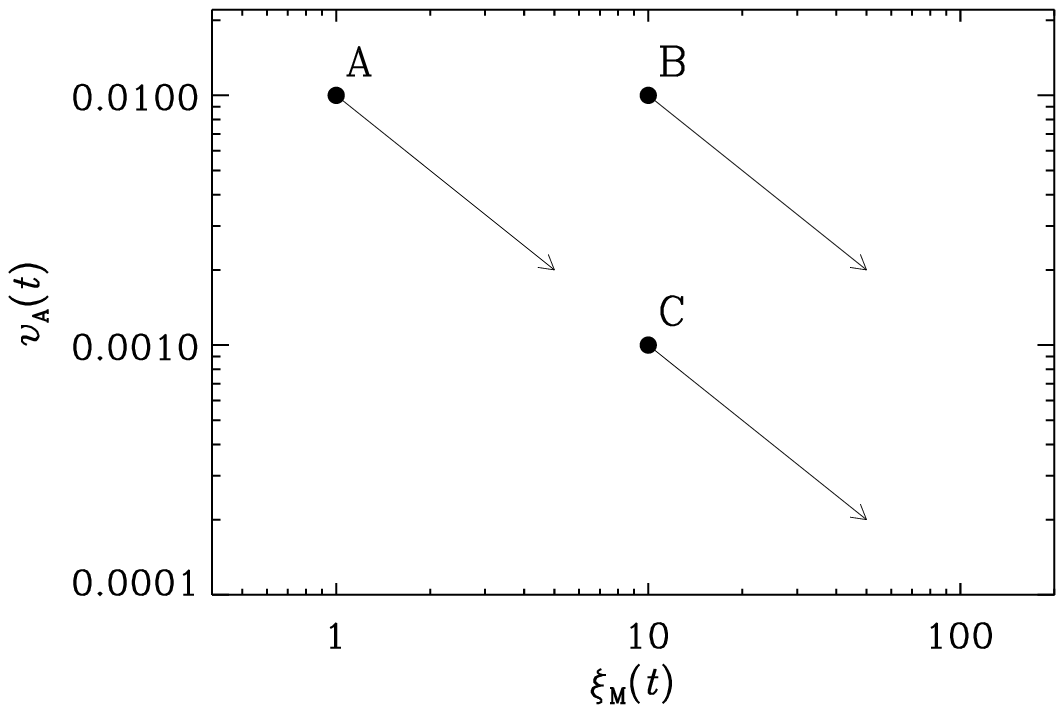}
\end{center}\caption[]{
Sketch of the three initial values of $\vA$ and $\xiM$ and their anticipated tracks.
}\label{psketch}\end{figure}

We perform runs such that their initial values of $\vA$ and $\xiM$ lie on three points as sketched in \Fig{psketch}.
Points A (Run A1) and B (Runs~B1, B2, and B3) have the same value of $\vA$ initially, but different values of $\xiM$,
while points B and C (Run C1) have the same value of $\xiM$, but different values of $\vA$.
Furthermore, the initial value of $\vA$ for Run~C1 is such that Run~A1 should meet up with it after some time.
Note also that points A and B, which start both at the nominal time $t=0$,
cannot lie on the same isochrone, if we expect the isochrone to lie on lines $\vA=\xiM/t_\ast$.
Since $I_\mathrm{A}\sim\vA^2\xiM^2$, points A and C have the same value of $I_\mathrm{A}$,
while point B has a 100 times larger value, because $\xiM^2$ is a hundred times larger.
The initial points A, B, and C could correspond to different generation mechanisms, or even to different model parameters for the same scenario.
Examples for the former are generation of magnetic fields during inflation or early Universe phase transitions (e.g., electroweak or QCD). 

We recall that in 2D decaying MHD turbulence, $\vA\propto t^{-1/2}$ and $\xiM\propto t^{1/2}$; see also \cite{BNV24}.
Therefore, the evolutionary track is $\vA\propto\xiM^{-1}$.
Thus, if point C has a 10 times larger value of $\xiM$ than point A, its value of $\vA$ should be 10 times smaller than for point A.
However, to account for a slightly faster decay at small Lundquist numbers, we choose correspondingly smaller initial value of $\vA$
in those cases (0.07 for Run~C1 and 0.09 for Runs~C4 and C4u).
Point~B is chosen such that it would lie on the same asymptotic isochrone as point~C, i.e., $\vA=\xiM/t_\ast$,
although the time $t_\ast$ of this isochrone cannot be the nominal time $t=0$ chosen in the simulation.

\begin{table*}[t]\caption{
Summary of runs with different starting positions given by $k_B$ and $B_0$.
For Run~C4u, the initial velocity is finite.
The anastrophy $I_\mathrm{A}$ was defined in \Sec{BasicEquations}.
The quantities $\CM$ and $t_\mathrm{offset}$ are defined in \Sec{AlfvenTime},
$t_\mathrm{proper}$ is defined in \Sec{EvolutionaryTracks}, while
$t_\xi$ and $t_\mathcal{E}$ are defined in \Sec{TemporalEvolution}.
}\vspace{12pt}\centerline{\begin{tabular}{lcccccccrrrrcc}
Run & $\Lu$ & $k_1$ & $k_B$ & $B_0$ & $u_0$ & $\eta=\nu$ & $\CM$ & $t_\mathrm{offset}$ & $t_\mathrm{proper}$ & $t_\xi\;$ & $t_\mathcal{E}\;$ & $I_\mathrm{A}/\vA^2\xiM^2$ & res. \\
\hline
A1 &   70 & 0.01  &   1 & 0.1  &  0   & $10^{-4}$        & 10.6 &   420 &   600 &   460 &   440 & 1.24 & $1024^2$ \\
B1 &   80 & 0.001 & 0.1 & 0.1  &  0   & $10^{-3}$        & 10.6 &  5700 &  6000 &  4600 &  4400 & 1.24 & $1024^2$ \\
C1 &   70 & 0.001 & 0.1 & 0.007&  0   & $10^{-4}$        &  9.4 & 70000 & 63000 & 57000 & 57000 & 1.23 & $1024^2$ \\
B2 &  190 & 0.001 & 0.1 & 0.1  &  0   & $5\times10^{-4}$ & 14   &  6100 &  5300 &  5100 &  5000 & 1.31 & $2048^2$ \\
B3 &  540 & 0.001 & 0.1 & 0.1  &  0   & $2\times10^{-4}$ & 19   &  8100 &  6600 &  6500 &  6000 & 1.37 & $4096^2$ \\
A4 & 1070 & 0.01  &   1 & 0.1  &  0   & $10^{-5}$        & 24   &  1100 &   740 &   810 &   740 & 1.41 & $8192^2$ \\
C4 & 1100 & 0.001 & 0.1 & 0.009&  0   & $10^{-5}$        & 22   & 99000 & 74000 & 64000 & 59000 & 1.39 & $8192^2$ \\
C4u& 1100 & 0.001 & 0.1 & 0.009& 0.003& $10^{-5}$        & 22   & 69000 & 76000 & 77000 & 71000 & 1.44 & $8192^2$ \\
%
%
\label{TSummary}\end{tabular}}\end{table*}

\subsection{Alfv\'en and decay times}
\label{AlfvenTime}

The instantaneous Alfv\'en time is defined as
\begin{equation}
t_\mathrm{A}(t)=\xiM(t)/\vA(t),
\end{equation}
where $\xiM$ was defined in \Sec{DiagnosticQuantities}.
However, it is in general not true that $t_\mathrm{A}(t)$ is proportional to $t$,
because it depends on an a priori arbitrary shift in the definition of time.
(Even the epoch of generation is subject to an arbitrary offset.)
We therefore define for each run this shift by computing a linear fit to $t_\mathrm{A}(t)$.
\begin{equation}
t_\mathrm{A}^\mathrm{fit}(t)=mt+t_0.
\label{fit}
\end{equation}
Here, $t_0$ is the run-specific shift and $m$ is a nondimensional parameter.
It would be unity---even at late times---if the decay time is just the Alfv\'en time, but in
general, the actual decay time is longer than $t_\mathrm{A}(t)$ by a factor $\CM$, as
discussed in the introduction.
It can then be defined as
\begin{equation}
\CM=1/m,
\label{CMresult}
\end{equation}
so that
\begin{equation}
t=\CM \, t_\mathrm{A}(t).
\end{equation}
The factor $\CM$ was already determined by \cite{BNV24} using a different method.
As discussed in the introduction, it was found to grow with the Lundquist number
and saturate near $\Lu\approx10^4$, independently of the value of $\Pm$.

Multiplying \Eq{fit} with $\CM$, we obtain
\begin{equation}
\CM t_\mathrm{A}^\mathrm{fit}(t)=t+\CM t_0=t+t_\mathrm{offset},
\label{offset}
\end{equation}
which shows that $\CM t_0\equiv t_\mathrm{offset}$ is the offset time by which
we would need to shift the time axis in order to match the proper time of the run.

\section{Results}

The main results of our simulations are given in \Tab{TSummary}, where we report the values of $\CM$
(which indicates how much longer the decay time is compared to the Alfv\'en time)
as obtained from \Eq{CMresult}, along with several other parameters, as explained below.
The values of $\CM$ are found to be in the range 10--20.
As expected, only for the runs starting at point B we have a different value of anastrophy ($I_\mathrm{A}$).
This is because of the 10 times larger value of the eddy magnetic scale, $\xiM$.
For Run~B1, the values of the viscosity and magnetic diffusivity have been increased by a factor of 10 relative to Run~A1, because the mesh spacing is here 10 times coarser.
Since the product $\eta k_B$ remained unchanged, the value of $\Lu$ is approximately the same.
On the other hand, for Runs~C1 and C4, $\vA$ is about 10 times smaller, and even though the mesh spacing is coarser,
$\eta$ and $\nu$ were chosen to be the same as for Runs~A1 and A4, respectively.

Run~C4u is similar to Run~C4, except that here the initial value of $u_0$ is nonzero.
Here we used $u_0/B_0\approx0.3$, which reproduces the velocity spectra from the end of Run~A (see below).
For the velocity, we take $k_u=0.5\,k_B$, $\alpha_\mathrm{sub}^{(u)}=2$, $\alpha_\mathrm{ine}^{(u)}=0.75$, and $K_{v}=8\,k_B$,
i.e., an exponential cutoff is used to model the end of the velocity inertial range.
By contrast, no such cutoff is used for the magnetic field, i.e., $K_B\to\infty$.

\begin{figure*}[t!]\begin{center}
\includegraphics[width=\textwidth]{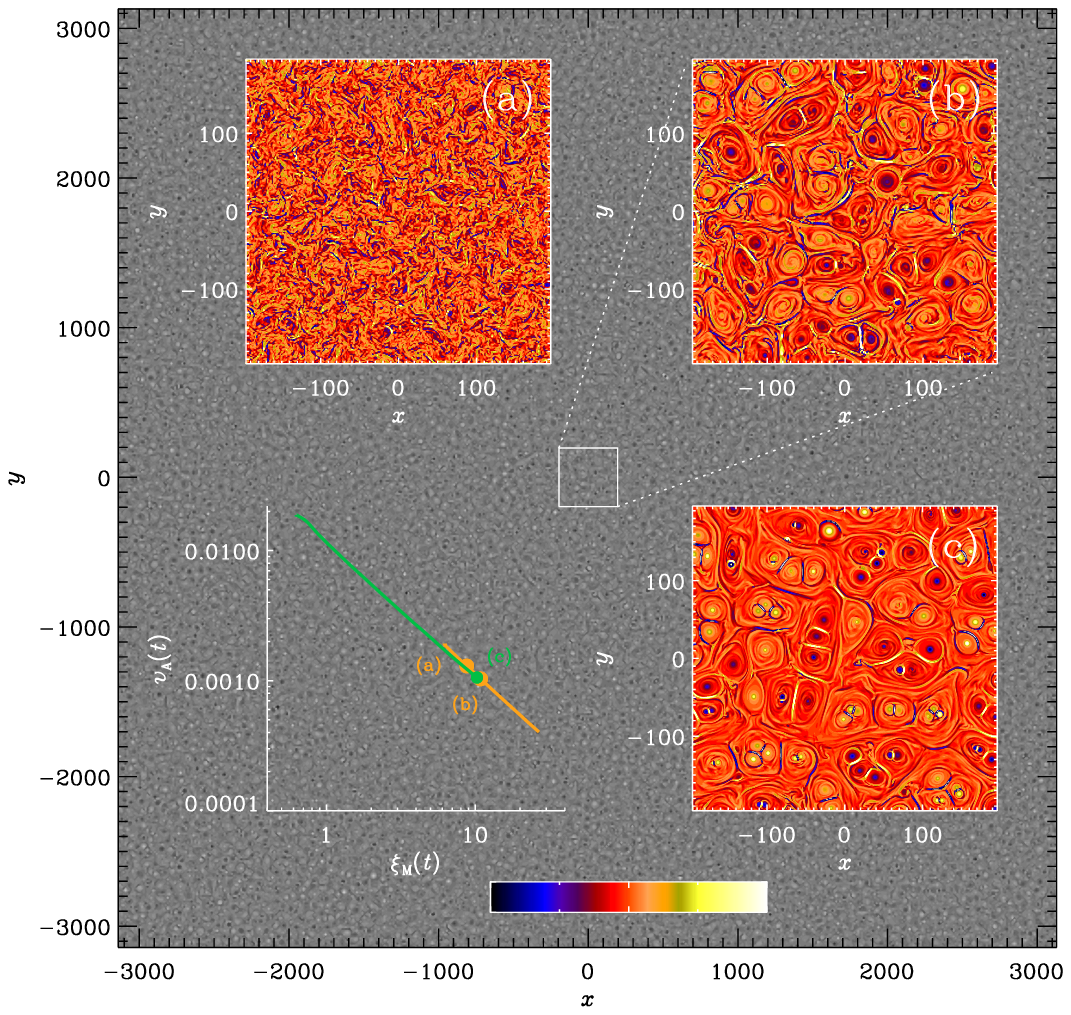}
\end{center}\caption[]{
Grayscale visualization of the current density along the line of sight, $J_z$, for Run~C4u at $t=10^5$.
The white square marks the inner part of the domain for which colorscale insets show
(a) Run~C4u at $t=5\times10^3$,
(b) Run~C4u at $t=10^5$, and
(c) Run~A4 at $t=2.3\times10^5$.
The corresponding positions in the $\xiM$ vs.\ $\vA$ diagram are shown in the lower left
inset for (a) and (b) on the track of Run~C4u in orange,
and for (c) on the track of Run~A4 in green.
The color table is qualitative and centered symmetrically around zero, corresponding to red shades,
with negative and positive extrema corresponding to blue and yellow shades, respectively.
The extrema vary between the three panels, so no numbers are given on the color bar.
}\label{pjz_comp}\end{figure*}

\begin{figure}[t!]\begin{center}
\includegraphics[width=\columnwidth]{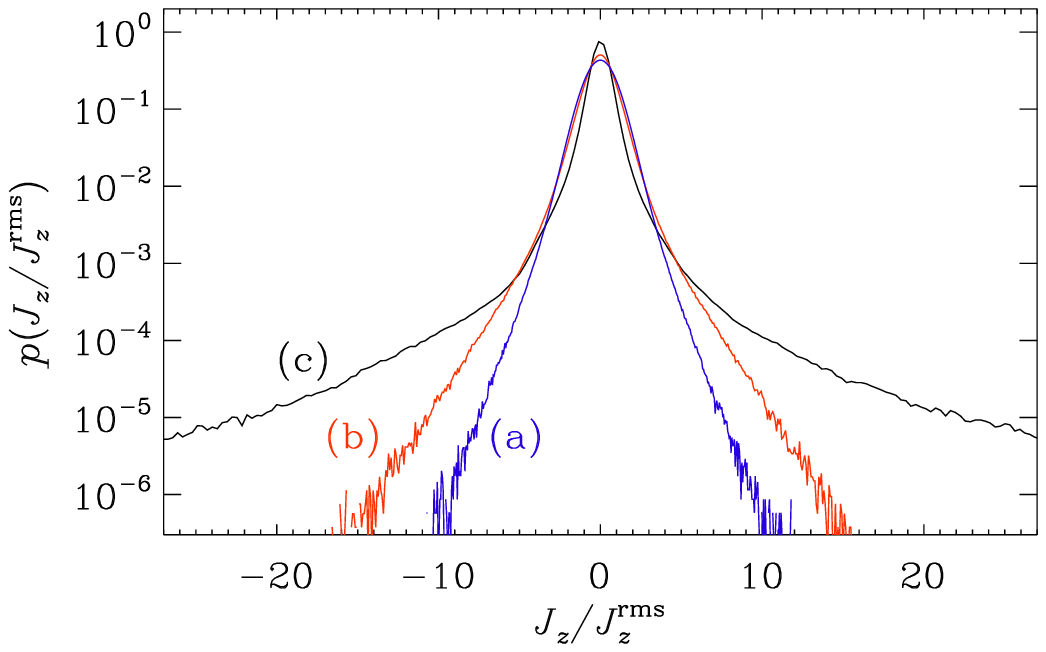}
\end{center}\caption[]{
Probability density functions $p(J_z/J_z^\mathrm{rms})$ for the three insets shown in \Fig{pjz_comp}.
}\label{pjz_hist_comp}\end{figure}

\begin{figure}[t!]\begin{center}
\includegraphics[width=\columnwidth]{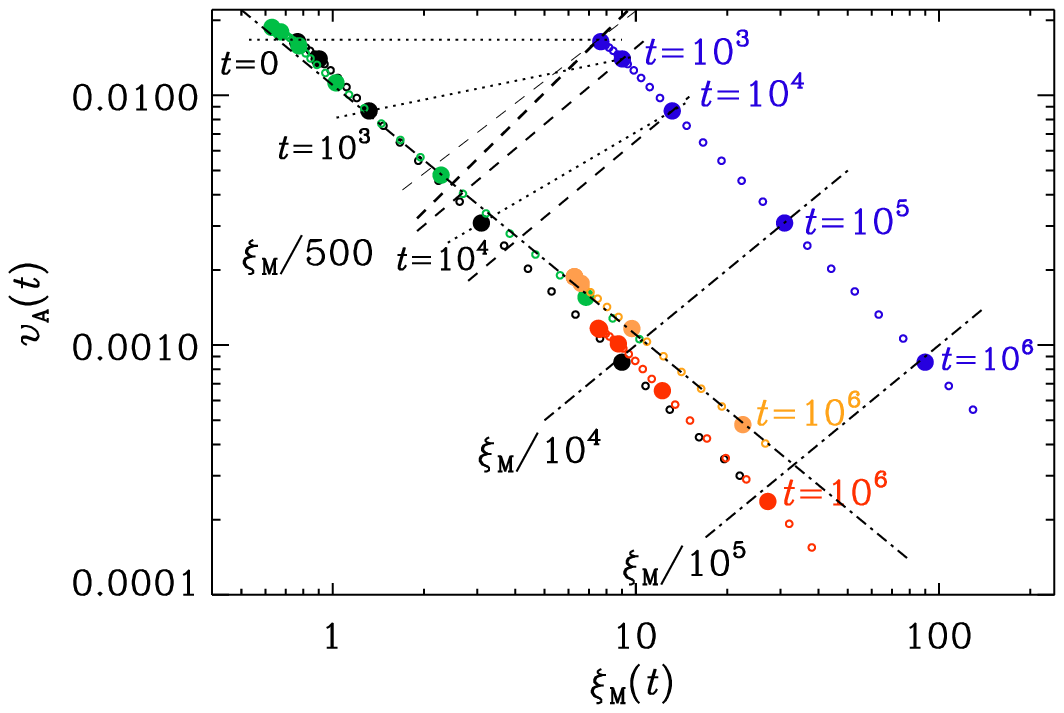}
\end{center}\caption[]{
Evolutionary tracks for Runs~A1 (black), B1 (blue), C1 (red), as well as Runs~A4 (green) and C4u (orange).
The symbols are logarithmically spaced in time with 6 open symbols per decade
and each decade is marked with a filled symbol.
The dashed-dotted lines are Alfv\'enic isochrones for $t_\ast=10^4$ and $t_\ast=10^5$.
The dotted lines are the \textit{nominal} isochrones for $t_\ast=0$, $t_\ast=10^3$, and $t_\ast=10^4$ between Runs~A1 and B1.
The dashed lines are the corrected isochrones (see text).
To demonstrate the effect of statistical errors, we show two dashed lines through the starting point for Run~B1 at $t=0$:
for the thicker one, a correction with $t_\mathrm{proper}=6000$ is applied, while for the thinner one $t_\mathrm{proper}=4400$ is used.
}\label{pkpm8192a}\end{figure}

\subsection{Visualizations}
\label{Visualizations}

We begin with visualizations of the $z$ component of the current density $J_z$ at the end of Run~A4 and the beginning of C4u at the approximate time $t=10^5$,
when the eddy scales of Run~C4u in the bigger domain agree with those of Run~A4 in the smaller domain; see \Fig{pjz_comp}.
For Run~C4u, we also show as an inset the result at $t=5\times10^3$, which is when the turbulent eddies are not yet fully developed.
The other two insets show the results for Run~A4, which is for a smaller domain, and Run~C4u.
In all cases, the insets show a zoomed part of the full domain and have the same size.

The images of $J_z$ shown in panels (a)--(c) of \Fig{pjz_comp} look rather different although the structures roughly agree in size and proper time.
One reason is that our initial conditions have random phases, while for truly turbulent flows, the phases are always correlated with the rest of the field;
compare the differences between panels (a) and (b).
Another reason is that $J_z$ develops strong departures from a Gaussian distribution.
This is demonstrated in \Fig{pjz_hist_comp}, where we show probability density functions
$p(J_z/J_z^\mathrm{rms})$ for the three insets shown in \Fig{pjz_comp}.
Note that for panels (b) and (c), $p(J_z/J_z^\mathrm{rms})$ develops stretched exponential tails.

\subsection{Evolutionary tracks}
\label{EvolutionaryTracks}

In \Fig{pkpm8192a}, we plot evolutionary tracks for Runs~A1--C1, as well as for Runs~A4 and C4u.
Here we also show isochrones and compare the nominal ones between Runs~A1 and B1 (dotted lines) with the corrected ones (dashed lines).
At early times, the nominal isochrones depart strongly from the asymptotic ones.
They are also not parallel to each other.
For example Runs~A1 and B1 have $t=0$ as the nominal initial time
and would therefore lie on a horizontal line.
Qualitatively, this can be explained by the nominal times not describing the ``proper'' time,
i.e., the time that ``belongs'' to a given value of $\xiM$ and $\vA$; namely
\begin{equation}
t_\mathrm{proper}=\CM\,\xiM(0)/\vA(0).
\label{tproper}
\end{equation}
Looking at \Fig{pkpm8192a}, the blue and black lines, belonging to Runs B1 and A1, respectively,
have the same value $\vA(0)$ ($\approx0.02$), but for Run B1,
$\xiM$ is 10 times larger than for Run~A1, and therefore $t_\mathrm{proper}$ is also 10 times larger; see \Eq{tproper}.
In order for this to be reflected in the time, we expect the shifted time $t+t_\mathrm{proper}$ for Run~B1 to be larger than that for~Run~A1.
This is indeed the case, and we have $t_\mathrm{proper}=6000$ for Run~B1 and $t_\mathrm{proper}=600$ for Run~A1; see \Tab{TSummary}.

To draw a revised isochrone for the initial time of these runs, we must connect the initial point of Run~B1
(with $t_\mathrm{proper}=6000$) with the corresponding proper time (5400, which yields 6000 with $t_\mathrm{proper}=600$) for Run~A1.
Likewise, for the corrected isochrones for $t_\ast=10^3$ and $10^4$, we must connect these points of Run~B1
with those of Run~A1 at $t=6400$ (which yields $7000=6000+10^3$ after adding $t_\mathrm{proper}=600$ for Run~A1)
and 15400 (which yields $16000=6000+10^4$ after adding 600), respectively.
This is done in \Fig{pkpm8192a}; see the dashed lines.
We see that the corrected isochrones are now nearly parallel to those at later times.
The isochrone for $t_\ast=0$ is actually slightly too steep (see the thicker one of the two dashed lines through $t=0$ for Run~B1),
but this could be explained by statistical errors, predominantly for $t_\mathrm{proper}$ of Run~B1.
To demonstrate this, we also show as the thinner dashed line the result for the lowest estimate of $t_\mathrm{proper}$ in \Tab{TSummary},
namely $t_\mathcal{E}=4400$, which yields a line that is now nearly perfectly parallel to the other lines.

We recall that $t_\mathrm{proper}$ has here been determined based on the initial values of $\CM t_\mathrm{A}(t)$.
As we have already argued above, it is therefore prone to errors.
By contrast, the quantity $t_\mathrm{offset}$ is based on the vertical offset $t_0$ of the fit given by \Eq{fit} in the graph $t_\mathrm{A}$ vs.\ $t$
within a given time interval, as will be discussed next.

\begin{figure}[t!]\begin{center}
\includegraphics[width=\columnwidth]{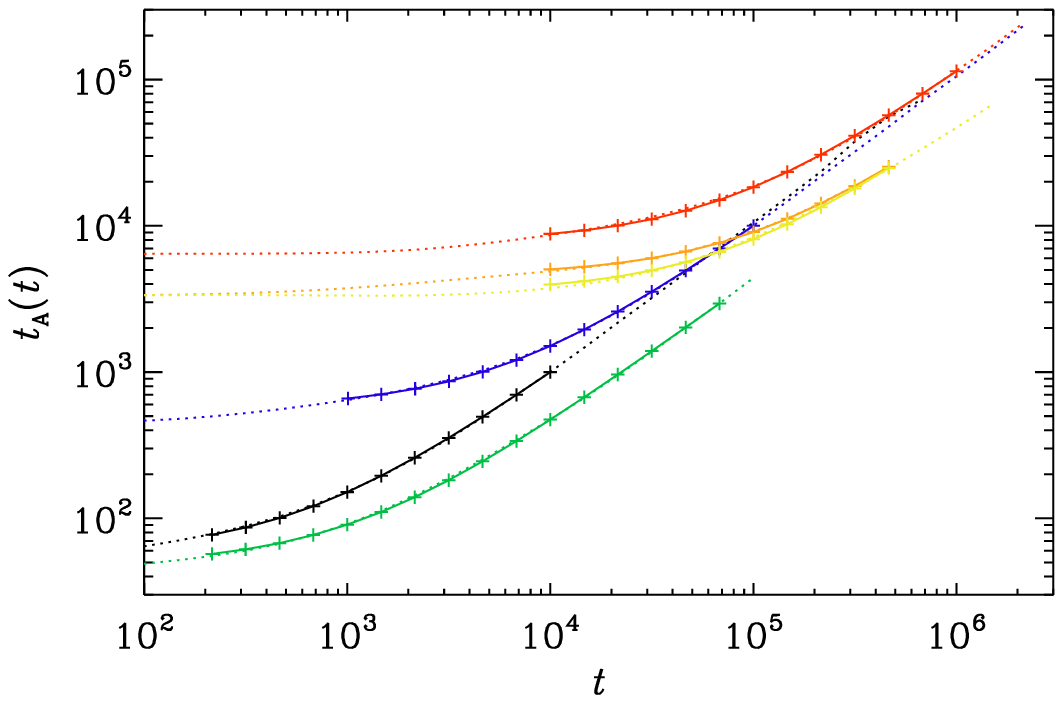}
\end{center}\caption[]{
Linear fits to $t_\mathrm{A}$ vs.\ $t$, giving $\CM=1/m$ and $t_0$ in \Eqs{fit}{CMresult}, as noted in \Tab{TSummary}.
The color coding is the same as in \Fig{pkpm8192a}.
The fits are shown as solid lines while the original data are dotted.
In this double-logarithmic representation, the linear fits appear curved.
The horizontal dotted lines indicate the extrapolation to the asymptote with the filled symbols marking the
values of $t_\xi$ and $t_\mathcal{E}$ given in \Tab{TSummary}.
}\label{pfit_8192_k1em3_k1em1_vA9em3_nu1em5u2}\end{figure}

\subsection{Fit to the Alfv\'en time}
\label{FitToAlfvenTime}

In \Sec{EvolutionaryTracks}, we discussed the usefulness of the quantity $t_\mathrm{proper}$ as a way of defining a proper initial time.
Another possibility is to determine the time $t_0$ from \Eq{fit}, as defined in \Sec{AlfvenTime}.
The result is shown in \Fig{pfit_8192_k1em3_k1em1_vA9em3_nu1em5u2}, where we present fits for Runs~A1, B1, C1, A4, C4, and C4u.
A difficulty here lies in the choice of a suitable range over which the fit should be applied.
The resulting parameters, which are listed in \Tab{TSummary}, can change somewhat, depending on the start and end time of the fit.
Nevertheless, the resulting values of $t_\mathrm{offset}=\CM t_0$ defined in \Eq{offset} agree reasonably well with the proper initial time
$t_\mathrm{proper}$ defined in \Sec{EvolutionaryTracks}, which here was based solely on the values of $\xiM$ and $\vA$ at the start of the simulation.
The differences in the results from the two approaches are therefore not surprising.

\begin{figure}[t!]\begin{center}
\includegraphics[width=\columnwidth]{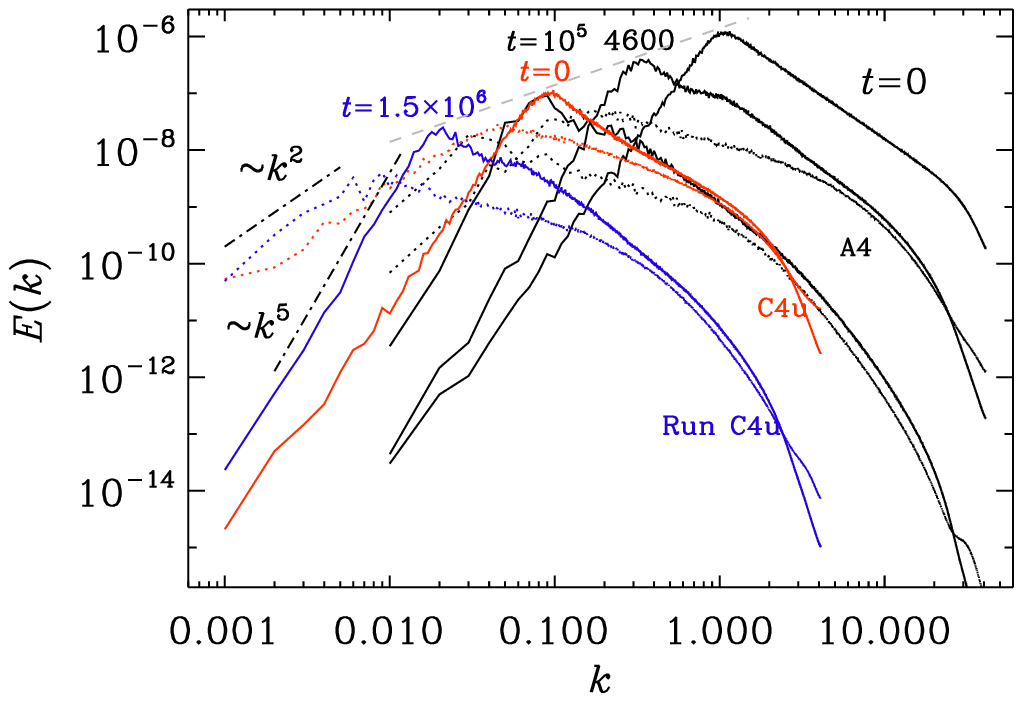}
\end{center}\caption[]{
Magnetic (solid lines) and kinetic (dotted lines) energy spectra for Runs~A4 and C4u.
Note that Run~C4u, which is for a ten times larger domain, connects smoothly to the end of Run~A4.
The gray dashed line indicates the expected envelope proportional to $k^\beta$ with $\beta=1$.
}\label{ppower_comp}\end{figure}

\subsection{Spectral evolution}

In \Fig{ppower_comp}, we plot spectra from Runs~A4 and C4u at different times.
The general evolution follows the expected behavior for magnetically dominated decaying turbulence \citep{Reppin+Banerjee17, 2017PhRvD..96l3528B, BNV24}.
As time goes on, the spectra shift in an approximately shape-invariant fashion to the lower left.
At high wavenumbers $k$, the spectral energy decreases, while instead it decreases at small wavenumbers, which is a clear signature of the occurrence of an inverse cascade.
The spectra evolve underneath an envelope $k^\beta$ with $\beta=1$; see \cite{BNV24}.

We also see that the initial spectrum of Run~C4u at $t=0$ approximately agrees with that of Run~A4 at $t=10^5$.
This suggests that the start time of Run~C4u should be shifted by a time span of the order $t=10^5$.
\Tab{TSummary} shows that this agrees with our values obtained using two different methods.

\subsection{Temporal evolution}
\label{TemporalEvolution}

The theoretically anticipated decay laws for $\xiM$ and $\EEM$ are
\begin{equation}
\xiM(t)\approx C_\xi I_\mathrm{A}^{1/4}\,\left(t_\xi+t\right)^{1/2},
\label{xiM}
\end{equation}
and
\begin{equation}
\vA(t)\approx C_{\vA} I_\mathrm{A}^{1/4}\,\left(t_\mathcal{E}+t\right)^{-1/2}.
\label{EEM}
\end{equation}
Qualitatively similar equations were also used by \cite{BNV24}, except that here we have added the offsets $t_\xi$ and $t_\mathcal{E}$.
The coefficients $C_\xi$ and $C_{\vA}$, from which we obtain the coefficient $C_\mathcal{E}=C_{\vA}^2/2$ for the magnetic energy,
are constants that have also been numerically determined in previous works.
More specifically, \cite{BNV24} found $C_\xi\approx0.13$ and $C_\mathcal{E}\approx15$ for their cases of 2D decay MHD,
and \cite{BYW25} have compared earlier results for the corresponding 3D cases,
where either the mean magnetic helicity density or the Hosking integral is conserved; see \App{2p5D_and_3D} for the corresponding decay laws.
A similarity of the coefficients across different applications might lend support to the idea that these coefficients are universal constant.

\begin{figure}[t!]\begin{center}
\includegraphics[width=\columnwidth]{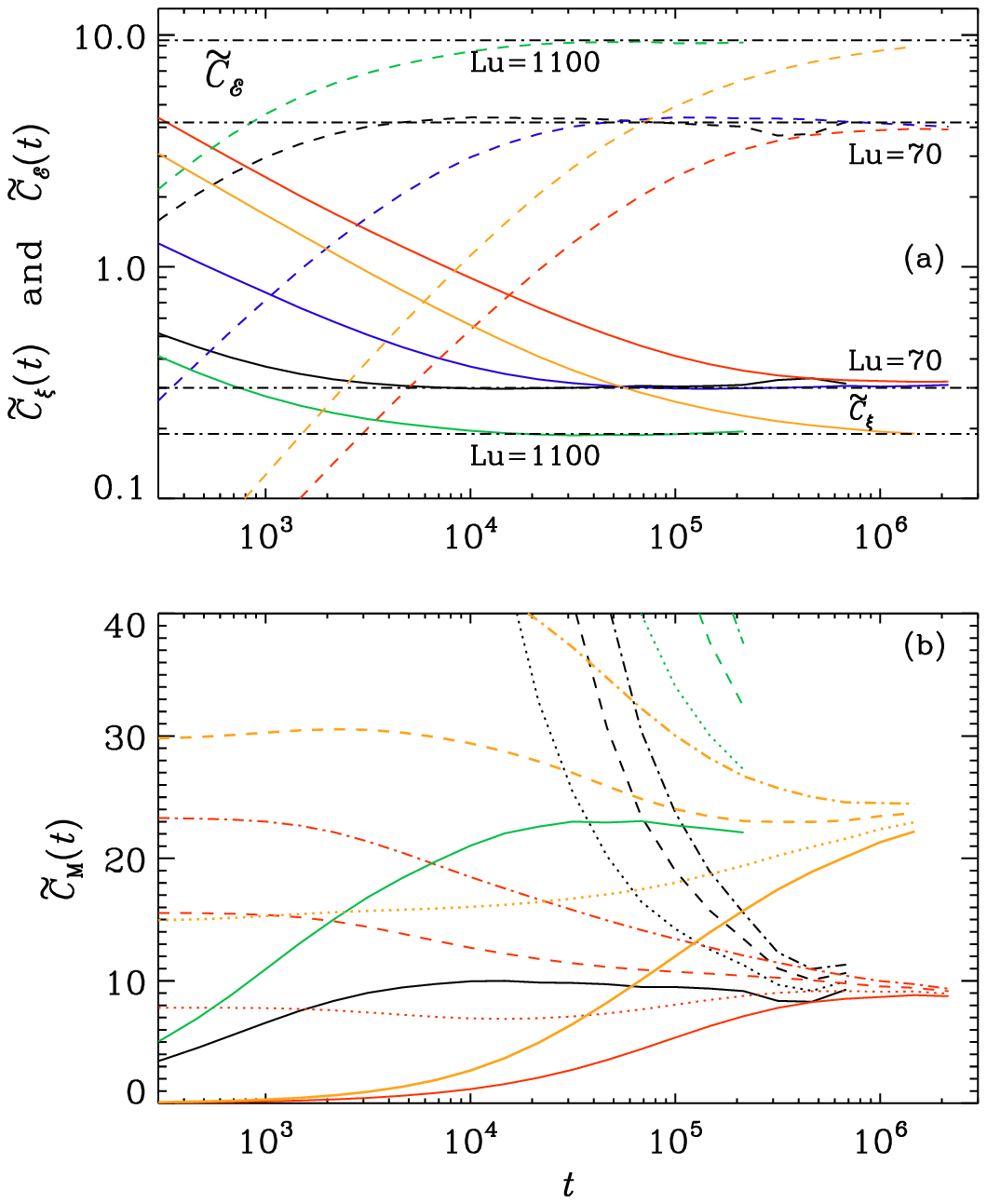}
\end{center}\caption[]{
(a) Dependence of $\tilde{C}_\xi(t)$ and $\tilde{C}_\mathcal{E}(t)$ for Runs~A1 (black), B1 (blue), C1 (red), as well as Runs~A4 (green) and C4 (orange).
(b) Dependence of $\tilde{C}_\mathrm{M}(t)$ for Runs~A1 (black), A4 (green), C1 (red), and C4 (orange).
Solid, dotted, dashed, and dashed-dotted lines refer to $\Delta t=0$, $5\times10^4$, $10^5$, and $1.5\times10^5$, respectively.
}\label{puniversal}\end{figure}

Independent estimates for $C_\xi$ and $C_\mathcal{E}$ can be obtained by plotting 
$\tilde{C}_\xi(t)=\xiM(t) \, I_\mathrm{A}^{-1/4}\,\left(t_\xi+t\right)^{-1/2}$ and
$\tilde{C}_\mathcal{E}(t)=\vA(t) \, I_\mathrm{A}^{-1/4}\,\left(t_\mathcal{E}+t\right)^{1/2}$,
and determining the height of their plateaus at late times; see \Fig{puniversal}(a).
The resulting values of $C_\xi$ and $C_\mathcal{E}$ fall into different groups for our small and large Lundquist number runs.
For small Lundquist numbers ($\Lu\approx70$), we have $C_\xi\approx0.3$ and $C_\mathcal{E}\approx4$, while for
larger values ($\Lu\approx1100$), we have $C_\xi\approx0.2$ and $C_\mathcal{E}\approx10$.
These values are larger than those reported in earlier work, but it is possible that the formerly reported values would be approached for larger Lundquist numbers.

At larger value of $\Lu$, also the coefficient $\CM$ in \Eqss{CMresult}{offset} increases.
It can also be obtained, similarly to $C_\xi$ and $C_\mathcal{E}$, as the plateau value of
\begin{equation}
\tilde{C}_\mathrm{M}(t;\Delta t)=(t+\Delta t)/t_\mathrm{A}(t),
\label{tildeCM}
\end{equation}
provided $\Delta t$ is close to the correct value.
These functions are shown in \Fig{puniversal}(b) for different values of $\Delta t$.
This plot provides not only an independent estimate of $\CM$ for large values of $t$, but it also shows how the evolution
of $\tilde{C}_\mathrm{M}(t;\Delta t)$ becomes approximately flat for a suitable value of $\Delta t$.
Comparing with \Eq{fit}, it is clear that $(t+\Delta t)/t_\mathrm{A}(t)$ should become equal to $\CM$, when $\Delta t=t_0/m=\CM t_0$.
This is similar to \Eq{offset}.
Computing $\tilde{C}_\mathrm{M}(t;\Delta t)$ therefore yields yet another method for computing both $\CM$ and $t_\mathrm{offset}$ at the same time.

The aforementioned method of extracting a plateau in the evolution of $(t+\Delta t)/t_\mathrm{A}(t)$ may provide a more robust technique in cases when the
time dependence of $t_\mathrm{A}(t)$ is more complicated and simple fits become problematic.
This may be the case when determining isochrones from actual magnetogenesis simulations.
In that case, one would not simply start from a static snapshot as initial condition,
but the magnetic field evolution would include its past history toward a particular point.
The function $\tilde{C}_\mathrm{M}(t;\Delta t)$ might then not yet have reached a plateau.
Similarly, at late times, this function may evolve away from the plateau value because of various computational artifacts.
So, again, that time stretch would need to be eliminated from the analysis.

\begin{figure}[t!]\begin{center}
\includegraphics[width=\columnwidth]{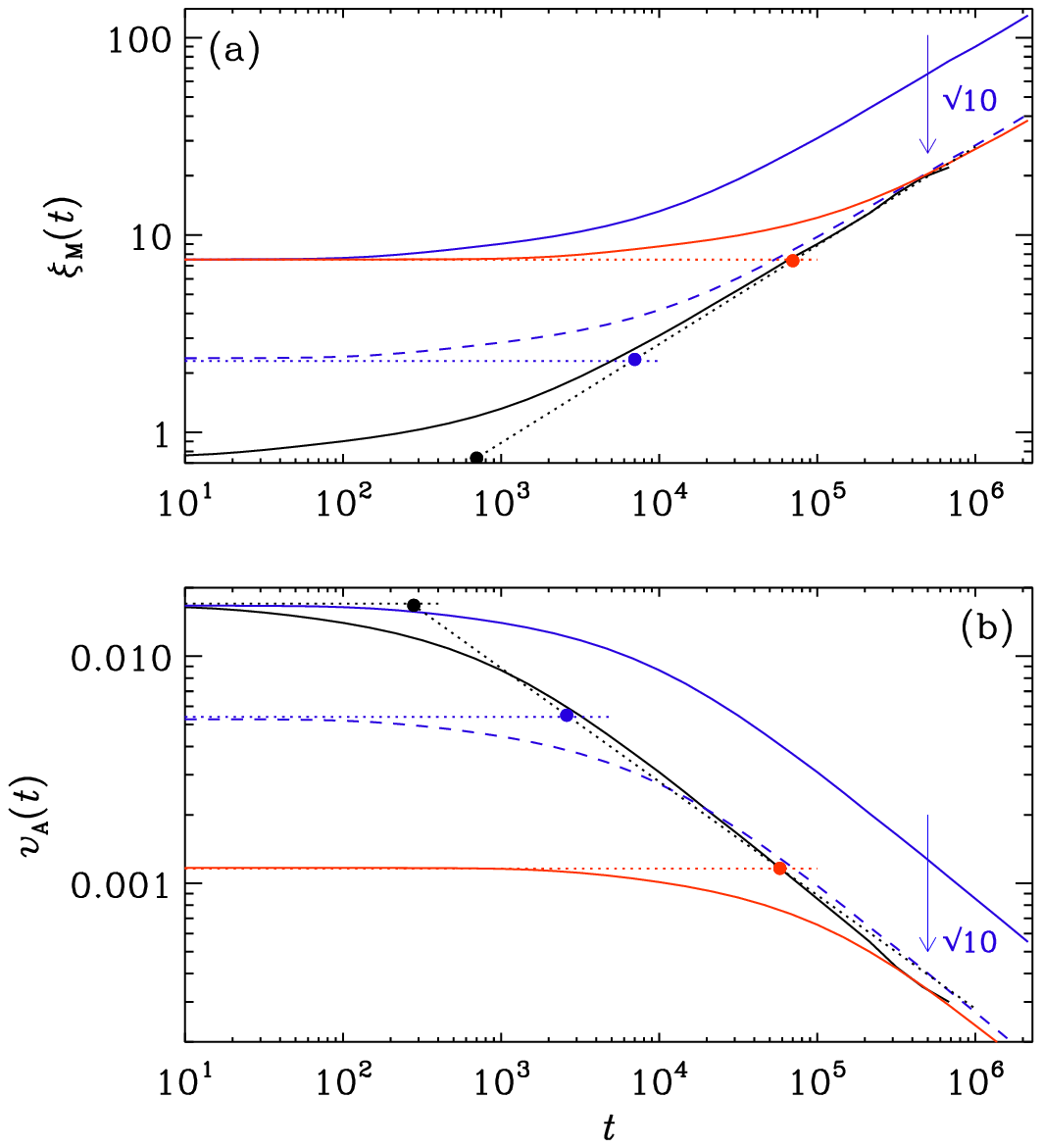}
\end{center}\caption[]{
Dependence of (a) $\xiM(t)$ and (b) $\vA(t)$ for Runs~A1 (black), B1 (blue), and C1 (red).
The final stages of Run~B1 (blue) can be collapsed onto those of Runs~A1 (black) and C1 (red).
}\label{pkpm1024a_t}\end{figure}

To verify the validity of \Eqs{xiM}{EEM}, we plot in \Fig{pkpm1024a_t} the evolution of $\xiM(t)$ and $\EEM(t)$ for Runs~A1, B1, and C1.
We see that, toward late times, the tracks of Runs~A1 and C1 agree approximately.
This is because these tracks also agree in \Fig{pkpm8192a}, where Run~C1 continues the evolution of Run~A1 toward larger scales.

In \Fig{pkpm1024a_t}(a), Run~B1 is shifted to the left while
in \Fig{pkpm1024a_t}(b) it is shifted to the right.
Therefore, Run~B1 cannot be made to agree with Runs~A1 or C1 by a shift in the time axis.
Instead, looking at \Fig{pkpm8192a}, we can collapse the blue line on top of the black and red lines
by a $45\degr$ shift along the isochrone to the lower left.
This corresponds to a combined shift by $1/\sqrt{10}$ both in $\vA$ and in $\xiM$.
Alternatively, we could have plotted $\xiM/I_\mathrm{A}^{1/4}$ and $\vA/I_\mathrm{A}^{1/4}$; see \Eqs{xiM}{EEM}.

In both panels of \Fig{pkpm1024a_t}, the horizontal dotted lines indicate the extrapolation to the asymptote
with the filled symbols marking the values of $t_\xi$ and $t_\mathcal{E}$.
For the values given in \Tab{TSummary}, we computed
$C_\xi=\min(\xiM/t^{1/2}I_\mathrm{A}^{1/4})$ and $t_\xi=\min(\xiM^2/C_\xi^2I_\mathrm{A}^{1/2})$, while
$C_{\vA}=\max(\vA t^{1/2}/I_\mathrm{A}^{1/4})$ and $t_{\vA}=\min(C_{\vA}^2I_\mathrm{A}^{1/2}/\vA^2)$.
The resulting values differ slightly from the intersection points indicated in \Fig{pkpm1024a_t}.

\subsection{Scaling exponents}

\begin{figure}[t!]\begin{center}
\includegraphics[width=\columnwidth]{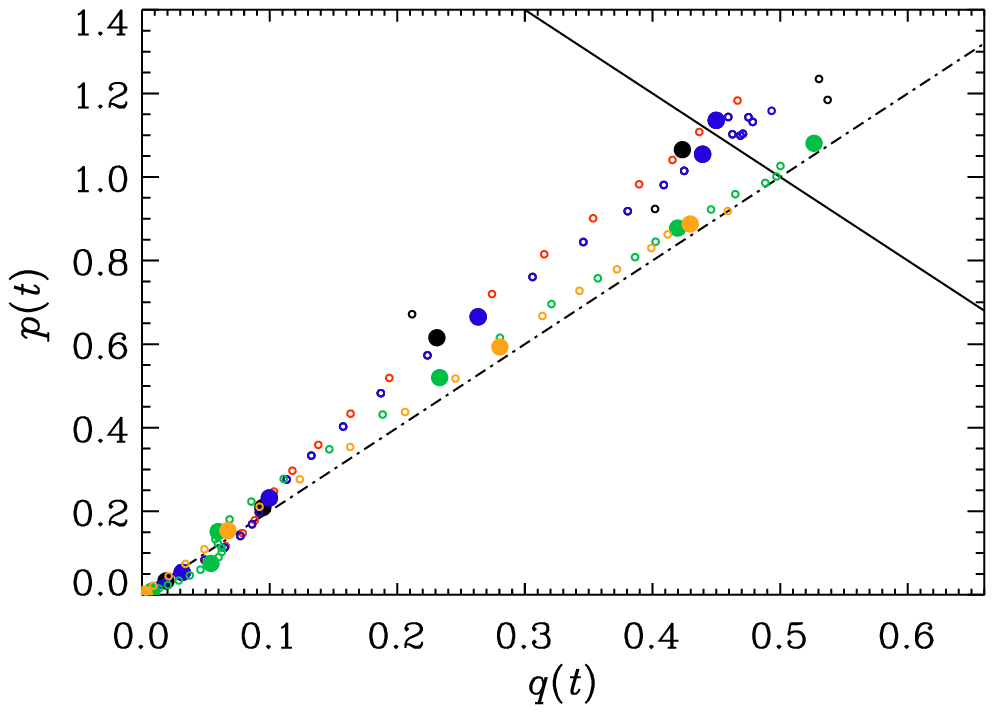}
\end{center}\caption[]{
Evolution in the $pq$ diagram for the same runs as in \Fig{pkpm8192a}, i.e., for Runs~A1 (black), B1 (blue), C1 (red), as well as Runs~A4 (green) and C4u (orange).
Again, the symbols are logarithmically spaced in time with 6 open symbols per decade and each decade is marked with a filled symbol.
The dashed-dotted line has the slope $\beta+1=2$ and the solid line corresponds to the Alfv\'en line where $p=2(1-q)$.
}\label{pkpm8192a_pq}\end{figure}

At limited Lundquist numbers and limited scale separation, there is always the possibility of departures from the theoretically expected values.
A robust way of inspecting the scaling properties of a run is then to plot the instantaneous scaling exponents
\begin{equation}
q(t)=\dd\ln\xiM/\dd\ln t,\quad
p(t)=-\dd\ln\EEM/\dd\ln t.
\end{equation}
Showing $p(t)$ vs.\ $q(t)$ tends to reveal an evolution along lines of constant $\beta(t)=p(t)/q(t)-1$, see \cite{BK17}.
In \Fig{pkpm8192a_pq}, we show such a plot for Runs~A1--C1, as well as Runs~A4 and C4u.

We see that, while the low Lundquist number runs still quite strongly deviate from the $\beta=1$ line,
the runs with larger values of $\Lu$ are much closer to that line.
All runs start in the lower left corner at $p=q=0$, and then settle around the Alfv\'en line where $q=1/2$ and $p=1$.
This is analogous to earlier findings for 3D turbulence \citep{BK17}.

\subsection{Conservation of anastrophy}

Given that the Lundquist number is finite, the anastrophy will not be perfectly conserved.
To quantify the degree of anastrophy conservation, we determine the decay exponent in $I_\mathrm{A}$ with time.
For a given Lundquist number, we find that $I_\mathrm{A}\propto t^{-s}$.
We have determined the value of $s$ for Runs~A1, B2, B3, and A4,
which cover the range $80\leq\Lu \leq 1100$.
We find that $s\ll1$, which suggests that the decay is slow.

\begin{figure}[t!]\begin{center}
\includegraphics[width=\columnwidth]{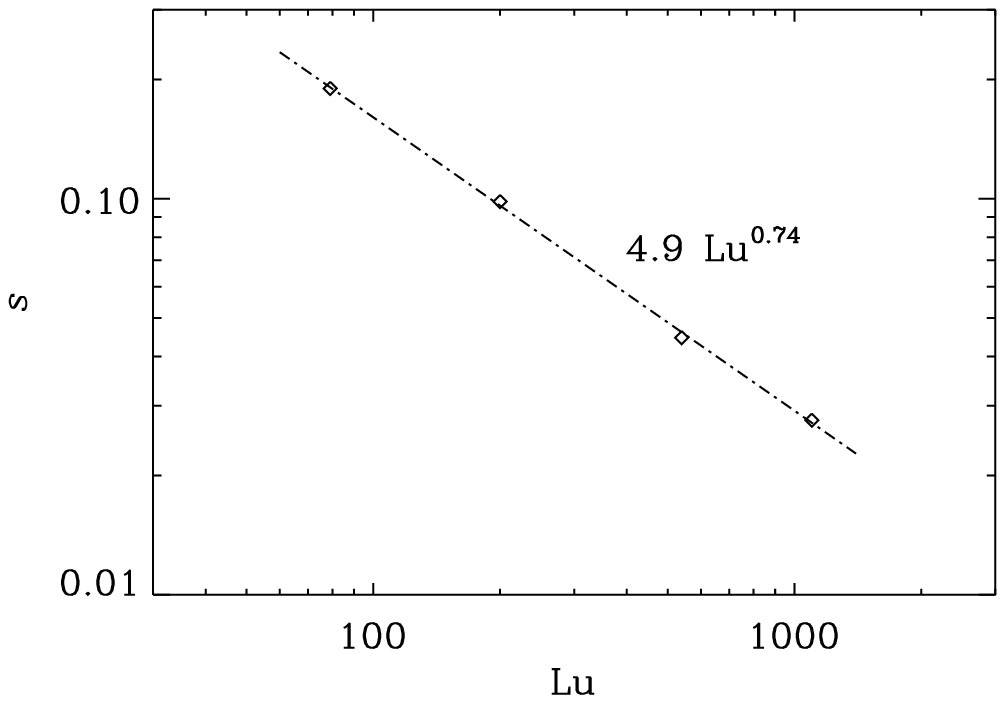}
\end{center}\caption[]{
Degree of anastrophy conservation quantified by the decay exponent $s$ as a function of $\Lu$.
}\label{panastrophy}\end{figure}

What is important, however, is the fact that the value of the exponent $s$ diminishes with increasing value of the Lundquist number.
The results are plotted in \Fig{panastrophy} and we see that $s\approx3.1\,\Lu^{-2/3}$.
Thus, for $\Lu=10^3$, we find $s=0.03$.
This would mean that the anastrophy would decay by a factor of 3 after 16 orders of magnitude in time, which is the range expected in cosmological applications.
Using $\Lu=10^6$, however, the anastrophy would decay by just 2 per cent after 30 orders of magnitude in time.

A similar analysis was done previously for the Hosking integral \citep{Zhou+22},
and it was found that the decay exponent decreases with Lundquist number approximately like $0.8\,\Lu^{-1/4}$.
To the best of our knowledge, similar decay laws have not previously been discussed.
It is therefore still unclear, how general our results are.

\subsection{Dependence on the initial velocity field}

\begin{figure}[t]\begin{center}
\includegraphics[width=\columnwidth]{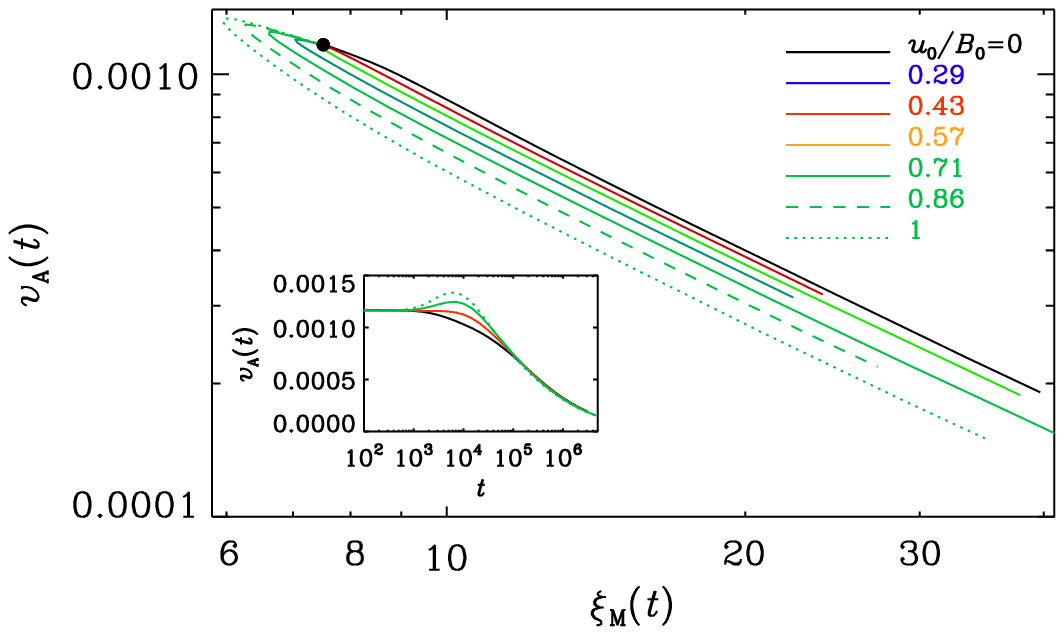}
\end{center}\caption[]{
Dependence of $\vA(t)$ vs.\ $\xiM(t)$ for different values of $u_0/B_0$.
The starting point is indicated by a black filled symbol.
The black line is for $u_0=0$ and shows a slight upward bend, while the blue line
for $u_0/B_0=0.29$ is nearly straight.
For larger values of $u_0/B_0$, the track moves first to the upper left,
corresponding to a slight increase in field strength (shown in the inset) and a smaller typical length scale,
before entering the usual decay track at slightly smaller field strength.
The insect only shows the lines for $u_0/B_0=0$, 0.43, 0.71, and 1.
}\label{pkpm1024a_u}\end{figure}

We now examine the dependence of the evolutionary tracks on the presence of an initial velocity field.
In \Fig{pkpm1024a_u}, we compare models with different initial values of $u_0/B_0$.
For $u_0=0$, the tracks show a small upward trend in the $\xiM$ vs.\ $\vA$ diagram;
see the black line in \Fig{pkpm1024a_u}.
When $u_0/B_0\approx0.3$, the track is  straighter.
Finally, when $u_0/B_0>0.3$, the track evolves first to the left, i.e., the magnetic length scales initially decrease,
and later the track develops a downward trend, before entering the expected decay behavior.
This is qualitatively explained by the effect of turbulent diffusion exerted by the excessive strength of the flow
compared to the flow amplitude from the magnetic driving alone.

\begin{figure}[t]\begin{center}
\includegraphics[width=\columnwidth]{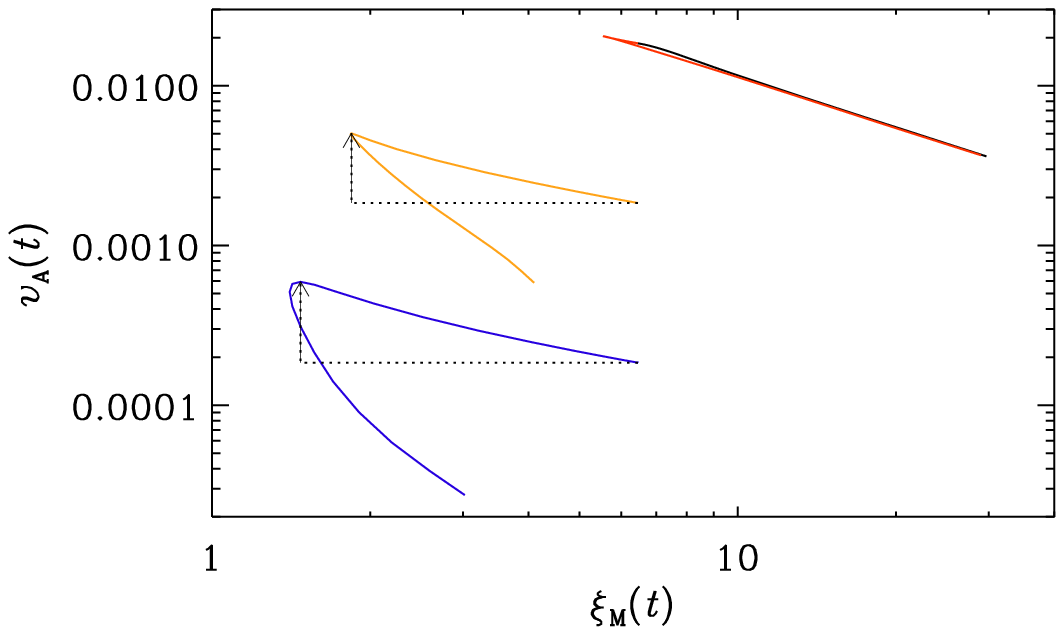}
\end{center}\caption[]{
Dependence of $\vA(t)$ vs.\ $\xiM(t)$ for Run~B3 with $u_0=0$ (black) and 0.05 (red), both with $B_0=0.1$,
as well as $B_0=10^{-2}$ (orange) and $B_0=10^{-3}$ again with $u_0=0.05$ (blue).
The vertical arrows indicate the increase of $\vA$ from the initial value to the maximum.
}\label{pkpm4096a_u}\end{figure}

It is possible that, for very large Lundquist numbers, a large velocity field not only enhances the decay of the magnetic field,
but that it actually leads to additional dynamo action and thereby to an enhancement of the magnetic field strength.
Dynamo action is only possible in 3D, but even in 2D, there can be some transient enhancement.
However, to see this more clearly, one would need to start with a magnetic field whose Alfv\'en velocity is smaller than the velocity field.
This is shown \Fig{pkpm4096a_u}, where we have lowered the magnetic field strength by a factor $100$.

The stronger magnetic field enhancement for weaker fields compared to stronger ones was seen quite clearly in the 3D decay simulations of \cite{BN25}.
Here, the magnetic field was growing during the early decay phase, when the magnetic Reynolds number based on the decaying velocity field was still large enough.
The growth was less pronounced when the initial magnetic field was stronger, which simply leads to an earlier saturation.
The simulations shown in \Fig{pkpm4096a_u} display a similar trend, even though in this case the growth is not caused by a dynamo.

\begin{figure}[t]\begin{center}
\includegraphics[width=\columnwidth]{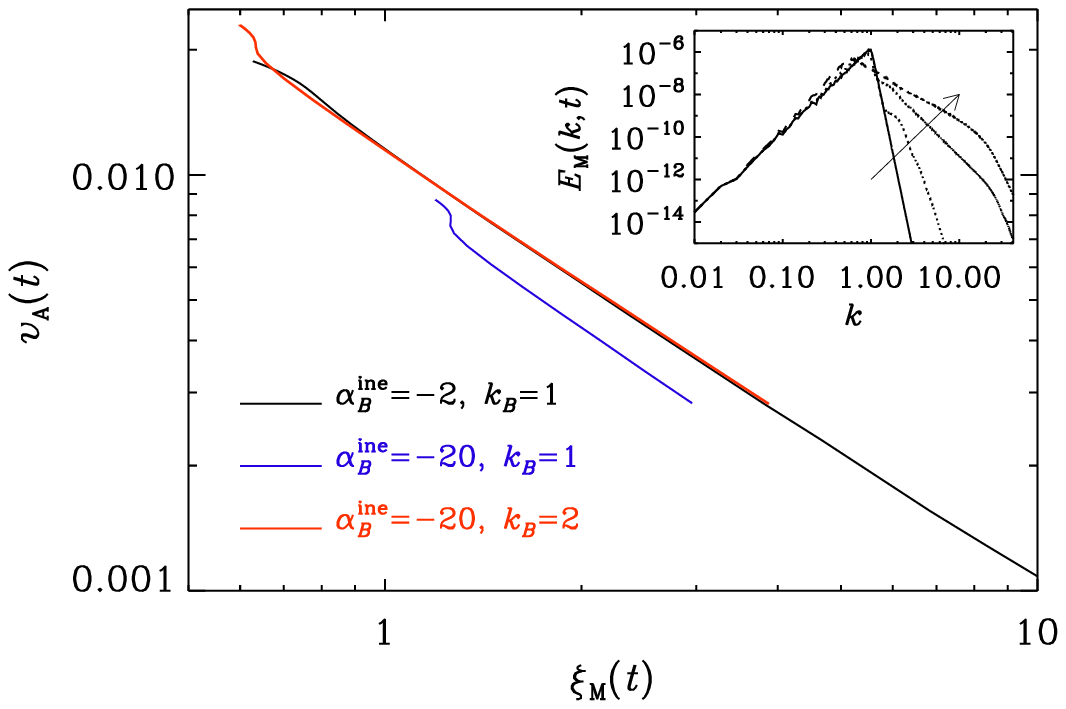}
\end{center}\caption[]{
Dependence of $\vA(t)$ vs.\ $\xiM(t)$ for Run~C4 with $\alpha^\mathrm{ine}_{B}=-2$ (the usual case, black) and
$\alpha^\mathrm{ine}_{B}=-20$ (red).
The inset shows the build-up of the inertial range at times $t=0$ (solid line),
$t=100$ and 300 (dotted lines), and $t=1000$ (dashed line),
corresponding to 0, 3, 8, and 11 Alfv\'en times.
}\label{pkpm8192a_u}\end{figure}

\subsection{Development of the inertial range}

In all the cases discussed above, we assumed that the initial magnetic field is described by a powerlaw spectrum
with an inertial range with a spectral slope $\alpha^\mathrm{ine}_{B}=-2$.
However, it is possible that the initial magnetic field had a very different spectrum, corresponding to a different magnetogenesis scenario.
In order to assess the effect of a mismatch between the initial spectrum and the spectrum that develops in the later self-similar phase,
we now consider a case with a very short inertial range.
By choosing $\alpha^\mathrm{ine}_{B}\to-\infty$, the spectrum would decay almost instantaneously with $k$.
In practice, we chose $\alpha^\mathrm{ine}_{B}=-20$, so that the inertial range is almost absent.
The comparison with the original Run~C4 is shown in \Fig{pkpm8192a_u}.
We see that with a significantly shorter inertial range, the length scale $\xiM$ is somewhat larger, and the initial field strength is slightly less.
Nevertheless, the initial values of $\xiM$ and $\vA$ still lie on a similar track as for the original Run~C4.
After a short time, however, an inertial range develops at the expense of the magnetic energy near the spectral peak.
This leads to a small drop of the field strength.
After that, however, the point $(\xiM,\vA)$ continues on a new track that corresponds to one with a slightly smaller anastrophy.
This is because the initial condition did already have a smaller initial value of the anastrophy.
This demonstrates the importance of characterizing the initial condition not just by their initial values of $\xiM$ and $\vA$,
but also by the value of the relevant conserved quantity.

To compensate for the larger initial length scale for runs with $\alpha^\mathrm{ine}_{B}=-20$,
we also compared in \Fig{pkpm8192a_u} with a run with $k_B=2$ instead of $k_B=1$.
Furthermore, to adjust for the resulting drop of the anastrophy, we increased the initial
field strength by a factor of 3.7.
In that case, the trajectory matches that for $\alpha^\mathrm{ine}_{B}=-2$ and $k_B=1$ reasonably well.
In \Fig{pkpm8192a_u}, we also show the build-up of the inertial range at subsequent times $t=100$, 300, and 1000,
corresponding to $t/t_\mathrm{A}\approx3$, 8, and 11 Alfv\'en times.
(The values of $t_\mathrm{A}=\xiM/\vA$ itself have slightly increased with time, because both $\xiM$ has increased and
$\vA$ has decreased, giving $t_\mathrm{A}=31$, 41, and 87 for these three times.)
Thus, we see that it takes at least ten Alfv\'en times to establish an inertial range.
This is consistent with recent requirements found by \cite{Silver+26} for magnetic field amplification in young galaxies.

\section{Conclusions}

A possibility to explain the origin of extragalactic magnetic fields is that volume-filling seed magnetic fields were generated
by primordial processes in the early Universe, e.g., during or after inflation, or during the electroweak or QCD phase transitions;
see \cite{sub15} and \cite{Vachaspati21} for recent reviews.
In each of these cases, it is crucial to understand how the magnetic field characteristics could have been modified by cosmic evolution,
in order to match their evolved configurations with possible constraints from observations of the local Universe \citep[e.g.,][]{2024arXiv241214825N,2025Univ...11..164C}.
The goal of this work was to derive universal isochrones which rule the evolution of primordial magnetic fields since their generation. 

As the primordial magnetic field decays due to the turbulence driven by the magnetic field itself,
its characteristic length scale increases in a predictable way until it reaches the line
$\vA\sim\xiM/t_\ast$ in the diagram of Alfv\'en speed vs.\ length scale \citep{BJ04}.
Here, $t_\ast$ is the time today, i.e., about $14\Gyr$.
This line is the final isochrone. 
At earlier times, i.e., for smaller values of $t_\ast$, we expect there to be similar isochrones that are all parallel to each other.
However, for very small values of $t_\ast$, it becomes important to know how exactly the time is defined.
Here, we have proposed different methods that give broadly similar results; see \Tab{TSummary}.

One might have thought that not all conceivable choices of initial field configurations are permissible,
and that their length scales must be compatible with what can be explained in terms of
the largest processed eddy that can develop in the time available.
This is, however, not the case.
Instead, any combination of initial values of $\xiM$ and $\vA$, or $k_{B}$ and $B_0$, is allowed.
All that is needed is to reset the time axis such that $t$ agrees with $\CM t_\mathrm{A}$.
This is demonstrated in \Sec{EvolutionaryTracks}, where we defined $t_\mathrm{proper}$ as a proper initial time of each simulation.
The other measures of $t_\mathrm{proper}$ can only be determined a posteriori after $\CM$ and $t_\mathrm{A}$ have been determined.
But this is not a restriction on the overall idea.

There is always the possibility of artifacts in simulating decaying turbulence that we should be kept in mind.
First of all, we have seen in \Sec{EvolutionaryTracks} that the expected evolutionary tracks are only obtained for large enough values of $\Lu$.
This also means that the resolution, i.e., the number of mesh points $N$, should be large enough.
However, there is an important compromise to strike between a small peak wavenumber $k_B$, which would increase the value of $\Lu$,
and a small value of $k_1$, which is needed to capture a sufficiently long subinertial scale range.
These factors can affect the determination of isochronous in unexpected ways.

There are several possible tools that could have been employed to optimize the use of the finite resolution available.
One is the use of time-dependent viscosities and magnetic diffusivities,
and another is the use of hyperviscosities and magnetic hyperdiffusivities \citep{HS21}.
Their effects have, to some extent, already been examined in \cite{Zhou+22}.
In the present work, these tools have not been used.
However, it would be interesting to see how the determination of isochrones would be affected when using these different tools to optimize the use of the resolution available.

In agreement with earlier findings, a given magnetic energy spectrum always implies a certain kinetic energy spectrum.
Starting with zero initial velocity always leads to an initial adjustment phase during which the velocity field reaches its
asymptotic selfsimilar state.
It is characterized by a peak at a wavenumber that is about half the wavenumber of the magnetic peak and a
hight that is about one third of the magnetic peak, followed by a spectrum with a slope $\sim k^{-3/4}$.
The reason for this empirically determined spectrum is still unclear.

The main purpose of our present work was to define proper offset times, such as $t_\mathrm{proper}$, $t_\mathrm{offset}$, $t_\xi$, $t_\mathcal{E}$, and $\Delta t$,
defined in \Secs{FitToAlfvenTime}{TemporalEvolution}.
There, we did already discuss the differences in the results from the five definitions.
The differences were relatively small and it is difficult to say, which of the five methods is the best one.
The ultimate test would be the application to early universe magnetogenesis mechanisms, when the magnetic field has just entered the radiation-dominated era.
In that case, the temporal evolution can be very complex so that a robust method must be applied.
This could well be the determination of $\Delta t$ from a plateau in the evolution of $\tilde{C}_\mathrm{M}(t;\Delta t)$.

\begin{acknowledgements}
We thank Chiara Caprini, Andrii Neronov, and Alberto Roper Pol for discussions and useful comments on an earlier version of the manuscript.
This research was supported in part by the European Research Council through the ERC Synergy Grant COSMOMAG under grant No.\ 101224803,
the Swedish Research Council (Vetenskapsr{\aa}det) under grant No.\ 2025-05957,
the National Science Foundation under grant Nos.\ NSF AST-2307698, AST-2408411, and NASA Award 80NSSC22K0825.
The work of OI is supported by the Swedish Research Council (Vetenskapsr{\aa}det) under the Starting Grant No.\ 2025-04140.
We acknowledge the allocation of computing resources provided by the
Swedish National Allocations Committee at the Center for Parallel Computers at the Royal Institute of Technology in Stockholm.
\\
\textbf{Software and data availability.} The source code used for
the simulations of this study, the {\sc Pencil Code} \citep{PC},
is freely available on \url{https://github.com/pencil-code/} with its latest developments.
The DOI of the code is https://doi.org/10.5281/zenodo.2315093.
The simulation setups and corresponding data are freely available on
\href{https://doi.org/10.5281/zenodo.20603413}{DOI:10.5281/zenodo.20603413}, as well as on
\href{https://norlx65.nordita.org/~brandenb/projects/Isochrones/}{norlx65.nordita.org/~brandenb/projects/Isochrones}.
\end{acknowledgements}


\appendix
\nolinenumbers
\section{Effect of initial subinertial range slope}
\label{SubinertialSlope}

\begin{figure}[h]\begin{center}
\includegraphics[width=\columnwidth]{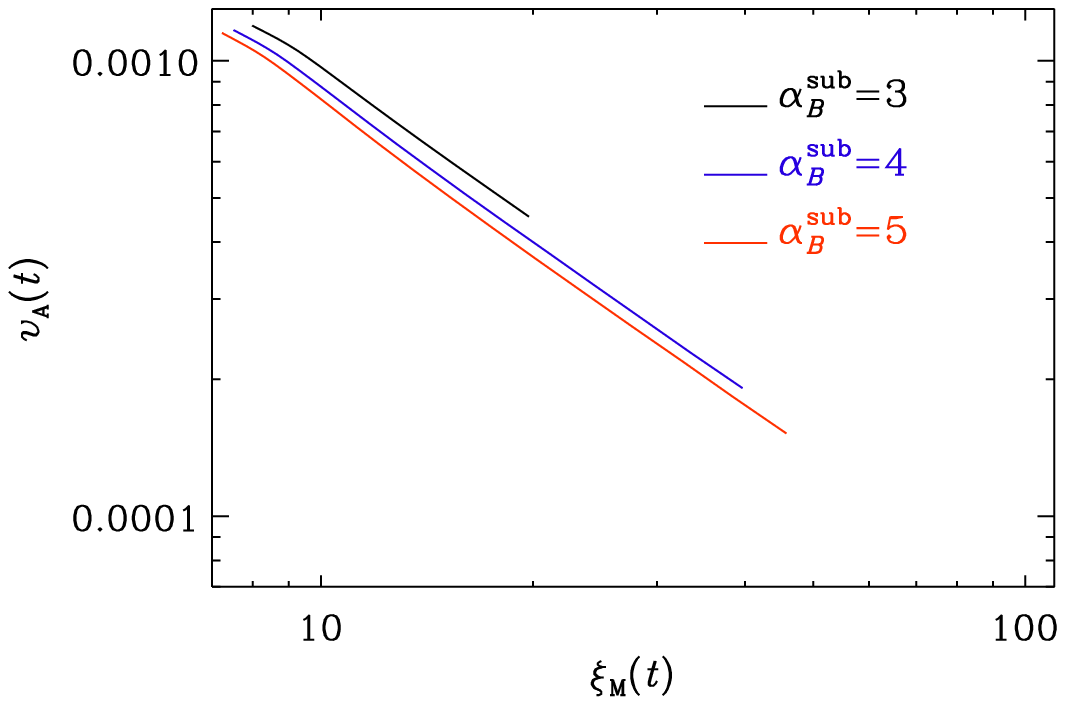}
\end{center}\caption[]{
Effect of different slopes $\alpha^\mathrm{sub}_{B}=3$, 4, and 5.
}\label{pkpm1024a_k}\end{figure}

As we have noted in \Sec{InitialConditions}, the subinertial range spectrum is not consistent with a causal one,
which would be $\propto k^3$ in two dimensions.
Instead, there is a trend to develop a steeper one.
In most of our runs, we have therefore used an initial $k^4$ spectrum for the magnetic field.
In this appendix, we show that the effect of the initial slope on the evolutionary track is negligible; see \Fig{pkpm1024a_k}
where we compare the tracks for $\alpha^\mathrm{sub}_{B}=3$, 4, and 5.

\begin{table}[h]\caption{
Connection between initial and final subinertial range slopes, compiled from BK17 \citep{BK17}, BSV23 \citep{BSV23}, and BK26 \citep{BK26}.
}\vspace{12pt}\centerline{\begin{tabular}{lccr}
Case & initial & final & reference \\
\hline
3D helical MHD  & $k^2$ & $k^4$ & BK17 \\
3D helical MHD  & $k^6$ & $k^4$ & BK17 \\
3D nonhelical MHD&$k^2$ & $k^2$ & BSV23 \\
2D Hall cascade & $k^4$ & $k^5$ & BK26 \\
\label{TSummary2}\end{tabular}}\end{table}

In \Tab{TSummary2}, we summarize some related results from earlier work.
In particular, for helical magnetic fields, \cite{BK17} found that an initial $k^2$ spectrum steepens to a $k^4$.
Conversely, a steeper initial $k^6$ becomes shallower, and is then again consistent with a $k^4$ spectrum.
On the other hand, for 3D nonhelical MHD, an initial $k^2$ spectrum remains $\propto k^2$ \citep{BSV23}.

During the time span when the magnetic field is still being generated, and it grows exponentially in time around a particular wavenumber $k_\ast$,
the spectrum for $k<k_\ast$ can have a slope steeper than $\propto k^4$.
An example is given by the chiral magnetic effect, where the magnetic energy spectrum was found to be $\propto k^{16}$; see Fig.~2 of \cite{Bra+17}.
This was argued to be caused by mode coupling, which affects the magnetic energy spectrum even at very small values of $k$.

The fact that the spectrum can instantaneously develop magnetic energy at very small $k$ is a consequence of the Fourier transform:
the Fourier decomposition of any spatially compact field configuration can contain modes at low $k$ values.
This should not be confused with a superalfv\'enic or even superluminal evolution of magnetic fields at large scales.

\section{Decay laws in 3D}
\label{2p5D_and_3D}

In \Sec{TemporalEvolution}, we presented the decay laws for $\xiM$ and $\EEM$ for the 2D case when the anastrophy is conserved.
This is not applicable to 3D, when the local magnetic helicity density is nonvanishing (even if it vanishes on average).
In that case, the Hosking integral $I_\mathrm{H}$ is conserved.
It characterizes the magnetic helicity fluctuations \citep{HS21, Zhou+22} and has dimensions of $\cm^{9}\s^{-4}$.
Therefore, instead of \Eqs{xiM}{EEM}, we now have
\begin{equation}
\xiM(t)\approx C_\xi^\mathrm{H} I_\mathrm{H}^{1/9}\,\left(t_\xi^\mathrm{H}+t\right)^{4/9},
\label{xiMH}
\end{equation}
and
\begin{equation}
\vA(t)\approx C_{\vA}^\mathrm{H} I_\mathrm{H}^{1/9}\,\left(t_\mathcal{E}^\mathrm{H}+t\right)^{-5/9},
\label{EEMH}
\end{equation}
where $t_\xi^\mathrm{H}$ and $t_\mathcal{E}^\mathrm{H}$ are the corresponding offsets.
This should be compared with the case when the average of the magnetic helicity density, $I_\mathrm{M}$, is nonvanishing.
It has dimensions of $\cm^{3}\s^{-2}$ and in that case, we have
\begin{equation}
\xiM(t)\approx C_\xi^\mathrm{M} I_\mathrm{M}^{1/3}\,\left(t_\xi^\mathrm{M}+t\right)^{2/3},
\label{xiMM}
\end{equation}
and
\begin{equation}
\vA(t)\approx C_{\vA}^\mathrm{M} I_\mathrm{M}^{1/3}\,\left(t_\mathcal{E}^\mathrm{M}+t\right)^{-1/3},
\label{EEMM}
\end{equation}
where $t_\xi^\mathrm{M}$ and $t_\mathcal{E}^\mathrm{M}$ are the corresponding offsets.
The transition from one to the other decay law depends on the relative strengths of $I_\mathrm{H}$ and $I_\mathrm{M}$,
and the values of the nondimensional coefficients $C_\xi^\mathrm{H}\approx0.14$, $C_\xi^\mathrm{M}\approx0.12$, and $C_{\vA}^\mathrm{H}\approx C_{\vA}^\mathrm{M}\approx4.0$;
see Eqs.~(3.8) and (3.9) of \cite{BB25}.

\bibliographystyle{aa}
\bibliography{refs.bib}

\begin{thebibliography}{47}
\expandafter\ifx\csname natexlab\endcsname\relax\def\natexlab#1{#1}\fi

\bibitem[{Banerjee \& Jedamzik(2004)}]{BJ04}
Banerjee, R. \& Jedamzik, K. 2004, \prd, 70, 123003

\bibitem[{{Bhat} {et~al.}(2021){Bhat}, {Zhou}, \& {Loureiro}}]{Bhat+21}
{Bhat}, P., {Zhou}, M., \& {Loureiro}, N.~F. 2021, \mnras, 501, 3074

\bibitem[{{Bondarenko} {et~al.}(2022){Bondarenko}, {Boyarsky}, {Korochkin},
  {Neronov}, {Semikoz}, \& {Sokolenko}}]{2022A&A...660A..80B}
{Bondarenko}, K., {Boyarsky}, A., {Korochkin}, A., {et~al.} 2022, \aap, 660,
  A80

\bibitem[{{Brandenburg}(2020)}]{Bra20}
{Brandenburg}, A. 2020, \apj, 901, 18

\bibitem[{{Brandenburg} \& {Banerjee}(2025)}]{BB25}
{Brandenburg}, A. \& {Banerjee}, A. 2025, J. Plasma Phys., 91, E5

\bibitem[{{Brandenburg} {et~al.}(1996){Brandenburg}, {Enqvist}, \&
  {Olesen}}]{BEO96}
{Brandenburg}, A., {Enqvist}, K., \& {Olesen}, P. 1996, \prd, 54, 1291

\bibitem[{{Brandenburg} \& {Kahniashvili}(2017)}]{BK17}
{Brandenburg}, A. \& {Kahniashvili}, T. 2017, \prl, 118, 055102

\bibitem[{{Brandenburg} {et~al.}(2017{\natexlab{a}}){Brandenburg},
  {Kahniashvili}, {Mandal}, {Pol}, {Tevzadze}, \&
  {Vachaspati}}]{2017PhRvD..96l3528B}
{Brandenburg}, A., {Kahniashvili}, T., {Mandal}, S., {et~al.}
  2017{\natexlab{a}}, \prd, 96, 123528

\bibitem[{{Brandenburg} \& {Kumar}(2026)}]{BK26}
{Brandenburg}, A. \& {Kumar}, V. 2026, JPlPh, to be submitted

\bibitem[{{Brandenburg} {et~al.}(2024){Brandenburg}, {Neronov}, \&
  {Vazza}}]{BNV24}
{Brandenburg}, A., {Neronov}, A., \& {Vazza}, F. 2024, \aap, 687, A186

\bibitem[{{Brandenburg} \& {Ntormousi}(2025)}]{BN25}
{Brandenburg}, A. \& {Ntormousi}, E. 2025, \apj, 990, 223

\bibitem[{{Brandenburg} {et~al.}(2017{\natexlab{b}}){Brandenburg}, {Schober},
  {Rogachevskii}, {Kahniashvili}, {Boyarsky}, {Fr{\"o}hlich}, {Ruchayskiy}, \&
  {Kleeorin}}]{Bra+17}
{Brandenburg}, A., {Schober}, J., {Rogachevskii}, I., {et~al.}
  2017{\natexlab{b}}, \apjl, 845, L21

\bibitem[{{Brandenburg} {et~al.}(2023){Brandenburg}, {Sharma}, \&
  {Vachaspati}}]{BSV23}
{Brandenburg}, A., {Sharma}, R., \& {Vachaspati}, T. 2023, JPlPh, 89, 905890606

\bibitem[{{Brandenburg} {et~al.}(2025){Brandenburg}, {Yi}, \& {Wu}}]{BYW25}
{Brandenburg}, A., {Yi}, L., \& {Wu}, X. 2025, JPlPh, 91, E113

\bibitem[{{Caprini} \& {Durrer}(2001)}]{Caprini+Durrer01}
{Caprini}, C. \& {Durrer}, R. 2001, \prd, 65, 023517

\bibitem[{{Caprini} \& {Durrer}(2006)}]{Caprini+Durrer06}
{Caprini}, C. \& {Durrer}, R. 2006, \prd, 74, 063521

\bibitem[{{Carretti} \& {Vazza}(2025)}]{2025Univ...11..164C}
{Carretti}, E. \& {Vazza}, F. 2025, Universe, 11, 164

\bibitem[{Christensson {et~al.}(2001)Christensson, Hindmarsh, \&
  Brandenburg}]{CHB01}
Christensson, M., Hindmarsh, M., \& Brandenburg, A. 2001, \pre, 64, 056405

\bibitem[{{Comisso} {et~al.}(2015){Comisso}, {Grasso}, \&
  {Waelbroeck}}]{Comisso+15}
{Comisso}, L., {Grasso}, D., \& {Waelbroeck}, F.~L. 2015, PhPl, 22, 042109

\bibitem[{{de Souza} \& {Opher}(2008)}]{2008PhRvD..77d3529D}
{de Souza}, R.~S. \& {Opher}, R. 2008, \prd, 77, 043529

\bibitem[{{Donnert} {et~al.}(2010){Donnert}, {Dolag}, {Brunetti}, {Cassano}, \&
  {Bonafede}}]{2010MNRAS.401...47D}
{Donnert}, J., {Dolag}, K., {Brunetti}, G., {Cassano}, R., \& {Bonafede}, A.
  2010, \mnras, 401, 47

\bibitem[{{Durrer} \& {Caprini}(2003)}]{DC03}
{Durrer}, R. \& {Caprini}, C. 2003, \jcap, 2003, 010

\bibitem[{{Durrer} {et~al.}(1998){Durrer}, {Kahniashvili}, \&
  {Yates}}]{Durrer+98}
{Durrer}, R., {Kahniashvili}, T., \& {Yates}, A. 1998, \prd, 58, 123004

\bibitem[{{Durrer} \& {Neronov}(2013)}]{2013A&ARv..21...62D}
{Durrer}, R. \& {Neronov}, A. 2013, \aapr, 21, 62

\bibitem[{{Frisch} {et~al.}(1975){Frisch}, {Pouquet}, {Leorat}, \&
  {Mazure}}]{Frisch+75}
{Frisch}, U., {Pouquet}, A., {Leorat}, J., \& {Mazure}, A. 1975, J. Fluid
  Mech., 68, 769

\bibitem[{{Fyfe} \& {Montgomery}(1976)}]{Fyfe+Montgomery76}
{Fyfe}, D. \& {Montgomery}, D. 1976, JPlPh, 16, 181

\bibitem[{{Ghosh} {et~al.}(2026){Ghosh}, {Brandenburg}, {Caprini}, {Neronov},
  \& {Vazza}}]{2026PhRvD.113b3523G}
{Ghosh}, O., {Brandenburg}, A., {Caprini}, C., {Neronov}, A., \& {Vazza}, F.
  2026, \prd, 113, 023523

\bibitem[{{Grasso} \& {Rubinstein}(2001)}]{GR01}
{Grasso}, D. \& {Rubinstein}, H.~R. 2001, \physrep, 348, 163

\bibitem[{{Hosking} \& {Schekochihin}(2021)}]{HS21}
{Hosking}, D.~N. \& {Schekochihin}, A.~A. 2021, PhRvX, 11, 041005

\bibitem[{{Hosking} \& {Schekochihin}(2023)}]{HS23}
{Hosking}, D.~N. \& {Schekochihin}, A.~A. 2023, NatCo, 14, 7523

\bibitem[{{Kahniashvili} {et~al.}(2008){Kahniashvili}, {Campanelli},
  {Gogoberidze}, {Maravin}, \& {Ratra}}]{Kahniashvili+08}
{Kahniashvili}, T., {Campanelli}, L., {Gogoberidze}, G., {Maravin}, Y., \&
  {Ratra}, B. 2008, \prd, 78, 123006

\bibitem[{{Kahniashvili} {et~al.}(2005){Kahniashvili}, {Gogoberidze}, \&
  {Ratra}}]{Kahniashvili+05}
{Kahniashvili}, T., {Gogoberidze}, G., \& {Ratra}, B. 2005, \prl, 95, 151301

\bibitem[{{Kahniashvili} {et~al.}(2013){Kahniashvili}, {Tevzadze},
  {Brandenburg}, \& {Neronov}}]{Kahn+13}
{Kahniashvili}, T., {Tevzadze}, A.~G., {Brandenburg}, A., \& {Neronov}, A.
  2013, \prd, 87, 083007

\bibitem[{{Kumar} \& {Brandenburg}(2026)}]{KB26}
{Kumar}, V. \& {Brandenburg}, A. 2026, JPlPh, submitted, arXiv:2605.18946

\bibitem[{{Neronov} {et~al.}(2024){Neronov}, {Vazza}, {Mtchedlidze}, \&
  {Carretti}}]{2024arXiv241214825N}
{Neronov}, A., {Vazza}, F., {Mtchedlidze}, S., \& {Carretti}, E. 2024, arXiv
  e-prints, arXiv:2412.14825

\bibitem[{{Pencil Code Collaboration} {et~al.}(2021){Pencil Code
  Collaboration}, {Brandenburg}, {Johansen}, {Bourdin}, {Dobler}, {Lyra},
  {Rheinhardt}, {Bingert}, {Haugen}, {Mee}, {Gent}, {Babkovskaia}, {Yang},
  {Heinemann}, {Dintrans}, {Mitra}, {Candelaresi}, {Warnecke},
  {K{\"a}pyl{\"a}}, {Schreiber}, {Chatterjee}, {K{\"a}pyl{\"a}}, {Li},
  {Kr{\"u}ger}, {Aarnes}, {Sarson}, {Oishi}, {Schober}, {Plasson}, {Sandin},
  {Karchniwy}, {Rodrigues}, {Hubbard}, {Guerrero}, {Snodin}, {Losada},
  {Pekkil{\"a}}, \& {Qian}}]{PC}
{Pencil Code Collaboration}, {Brandenburg}, A., {Johansen}, A., {et~al.} 2021,
  JOSS, 6, 2807

\bibitem[{{Pouquet}(1978)}]{Pouquet78}
{Pouquet}, A. 1978, JFM, 88, 1

\bibitem[{{Pouquet}(1993)}]{Pouquet93}
{Pouquet}, A. 1993, Les Houches Session XLVII, 139

\bibitem[{{Pouquet} {et~al.}(1976){Pouquet}, {Frisch}, \& {Leorat}}]{PFL76}
{Pouquet}, A., {Frisch}, U., \& {Leorat}, J. 1976, JFM, 77, 321

\bibitem[{{Reppin} \& {Banerjee}(2017)}]{Reppin+Banerjee17}
{Reppin}, J. \& {Banerjee}, R. 2017, \pre, 96, 053105

\bibitem[{{Silver} {et~al.}(2026){Silver}, {Lark}, \& {Schober}}]{Silver+26}
{Silver}, V., {Lark}, D., \& {Schober}, J. 2026, in prep.

\bibitem[{{Subramanian}(2016)}]{sub15}
{Subramanian}, K. 2016, Rep. Prog. Phys., 79, 076901

\bibitem[{{Tjemsland} {et~al.}(2024){Tjemsland}, {Meyer}, \& {Vazza}}]{tj24}
{Tjemsland}, J., {Meyer}, M., \& {Vazza}, F. 2024, \apj, 963, 135

\bibitem[{Vachaspati(2021)}]{Vachaspati21}
Vachaspati, T. 2021, Rept. Prog. Phys., 84, 074901

\bibitem[{Vazza {et~al.}(2025)Vazza, Gheller, Zanetti, Tsizh, Carretti,
  Mtchedlidze, \& Br\"{u}ggen}]{Vazza2025}
Vazza, F., Gheller, C., Zanetti, F., {et~al.} 2025, \aap, 696, A58

\bibitem[{{Widrow} {et~al.}(2012){Widrow}, {Ryu}, {Schleicher}, {Subramanian},
  {Tsagas}, \& {Treumann}}]{wi11}
{Widrow}, L.~M., {Ryu}, D., {Schleicher}, D.~R.~G., {et~al.} 2012, \ssr, 166,
  37

\bibitem[{{Zhou} {et~al.}(2022){Zhou}, {Sharma}, \& {Brandenburg}}]{Zhou+22}
{Zhou}, H., {Sharma}, R., \& {Brandenburg}, A. 2022, JPlPh., 88, 905880602

\end{thebibliography}
\end{document}